\documentclass[prd,aps,showpacs,epsf,floats,onecolumn]{revtex4}%
\usepackage{amssymb}
\usepackage{amsfonts}
\usepackage{amsmath}
\usepackage{graphicx}%
\providecommand{\U}[1]{\protect\rule{.1in}{.1in}}
\begin{document}
\title{\textbf{From Quantum-Mechanical Acceleration Limits to Upper Bounds on
Fluctuation Growth of Observables in Unitary Dynamics}}
\author{\textbf{Carlo Cafaro}$^{1}$, \textbf{Walid Redjem}$^{1}$, \textbf{Paul M.
Alsing}$^{1}$, \textbf{Newshaw Bahreyni}$^{2}$, \textbf{Christian Corda}$^{3}$}
\affiliation{$^{1}$University at Albany-SUNY, Albany, NY 12222, USA}
\affiliation{$^{2}$Pomona College, Claremont, CA 91711, USA}
\affiliation{$^{3}$SUNY Polytechnic Institute, Utica, NY 13502, USA}

\begin{abstract}
Recently, the notion of a quantum acceleration limit has been proposed for any
unitary time evolution of quantum systems governed by arbitrary nonstationary
Hamiltonians. This limit articulates that the rate of change over time of the
standard deviation of the Hamiltonian operator---representing the acceleration
of quantum evolution within projective Hilbert space---is constrained by the
standard deviation of the time-derivative of the Hamiltonian, expressed as
$\dot{\sigma}_{\mathrm{H}}\leq\sigma_{\mathrm{\dot{H}}}$. In this paper, we
extend our earlier findings to encompass any observable $A$ within the
framework of unitary quantum dynamics, leading to the inequality $\dot{\sigma
}_{A}\leq\sigma_{v_{A}}$. This relationship signifies that the speed of the
standard deviation of any observable is limited by the standard deviation of
its associated velocity-like observable $v_{A}$. Finally, for pedagogical
purposes, we illustrate the relevance of our inequality by providing clear
examples. We choose suitable observables related to the unitary dynamics of
two-level quantum systems, as well as a harmonic oscillator within a
finite-dimensional Fock space.

\end{abstract}

\maketitle

\section{Introduction}

Quantum Speed Limits (QSLs) study the time constraints on how fast the quantum
state of a system can evolve from an initial state to a final state.
Furthermore, they play an important role in quantum computation and quantum
information. The two main approaches to quantum speed limits were presented by
the Mandelstam-Tamm (MT) \cite{MT} bound and the Margolus-Levitin (ML)
\cite{ML} bound. A unified bound, resulting from the integration of these two
bounds, appears in Ref. \cite{LT}. The Mandelstam-Tamm (MT) bound sets a
fundamental limit on the speed of quantum evolution. In a closed system where
the evolution is governed by the Schr\"{o}dinger equation, it explains that
the time $\tau$ required for such a quantum system to evolve from an initial
state to a final state is limited by the system's energy uncertainty $\tau
\geq\pi\hslash/(2\Delta E)$, where the energy uncertainty $\Delta E$ is given
by $\Delta E\overset{\text{def}}{=}\sqrt{\langle\mathrm{H}^{2}\rangle
-\langle\mathrm{H}\rangle^{2}}$, with $\mathrm{H}$ being the generally
time-dependent Hamiltonian of the system \cite{MT}. For completeness, the
derivation of the MT bound is given in Appendix A. The Margolus-Levitin (ML)
bound is given by $\tau\geq\pi\hslash/(2\langle E\rangle)$ and is another
speed bound which involves the system's average energy $\langle E\rangle$
instead of its energy uncertainty. It describes how the total available energy
limits the evolution of a quantum system. The actual speed limit for the
quantum system is found by taking the maximum of the\textbf{ }MT and ML bounds
\cite{ML,LT,campo25}\textbf{,} $\tau\geq\max\left[  \pi\hslash/(2\Delta
E)\text{, }\pi\hslash/\left(  2\left\langle E\right\rangle \right)  \right]  $.

A generalization of QSLs to describe the dynamics of macroscopic observables
in large systems is investigated in \cite{H1}. It is shown that the speed
limit given by taking the maximum of the MT and ML bounds remains valid for
classical systems as well as quantum systems. Furthermore, a generalization of
QSLs applied to the dynamics of fluctuations in observables is considered in
\cite{H2}. Instead of the evolution between orthogonal quantum states, this
new limit can be used for systems experiencing fluctuations. Fluctuation
theorems \cite{SP,rondoni,campisi,landi}, widely used in nonequilibrium
systems, study the behavior of probability distributions of certain
observables under time-reversal symmetry. However, the evolution of
fluctuations over time has not been the focus of studies. Two primary
statistical measures employed to characterize the variable nature of quantum
system dynamics are the mean and the standard deviation of observables that
are directly pertinent to experimental assessments. While the dynamics of the
mean value has been explored for both closed and open quantum systems across
pure and mixed states \cite{MP,Pi}, the limitations on the speed of an
observable's fluctuation, or its standard deviation, have remained largely
unaddressed. Hamazaki demonstrated in Ref. \cite{H2} that, within the
framework of both unitary and certain dissipative quantum dynamics, the rate
of fluctuation of an observable is constrained by the fluctuation of a
relevant observable that represents velocity. Quantum speed limits are also
generalized for systems involving multiple observers \cite{H3}. In Ref.
\cite{H2}, universal bounds on the time-dependence of fluctuations for both
classical and quantum systems are studied based on an inequality indicating
that the standard deviation of any time-dependent observable $A$ has a speed
that is always smaller than the standard deviation $\sigma_{v_{A}}$ of the
suitably chosen velocity observable $v_{A}$, $\left\vert d\sigma
_{A}/dt\right\vert \leq\sigma_{v_{A}}$. In this inequality, $\sigma
_{A}\overset{\text{def}}{=}\sqrt{\langle A^{2}\rangle-\langle A\rangle^{2}}$
is the standard deviation of the observable $A$, $\left\langle A\right\rangle
$ is the average value of $A$, and the velocity observable denoted by $v_{A}$
is such that its expectation value is given by $d\langle A\rangle/dt=\langle
v_{A}\rangle$, with $\sigma_{v_{A}}$ being the standard deviation of the
velocity observable.

Quantum Acceleration Limits (QALs) are built upon QSLs and focus on how fast
the system's speed in Hilbert space can change. In Refs. \cite{P,AC}, an upper
limit was established for the rate of change in the speed of transportation
within any arbitrary finite-dimensional projective Hilbert space. This limit
subsequently introduced the concept of a quantum acceleration threshold
applicable to any unitary time evolution of quantum systems governed by
arbitrary nonstationary Hamiltonians. Time-dependent Hamiltonians are studied
in Ref. \cite{P} based on the Robertson-Schr\"{o}dinger uncertainty relation
\cite{schrodinger}. For systems evolving under such Hamiltonians, the rate of
change of speed of transportation in projective Hilbert space is defined as
acceleration where the speed is given by the square root of the variance in
the Hamiltonian of the system. It is shown that the acceleration $a$ in
projective Hilbert space is upper bounded as $a^{2}\leq(\Delta\mathrm{\dot{H}%
})^{2}/\hslash^{2}$ where $\Delta\mathrm{\dot{H}}$ is the uncertainty of
$\mathrm{\dot{H}}$ \cite{P}, with $\mathrm{\dot{H}}$ being the time derivative
of a nonstationary Hamiltonian operator $\mathrm{H}=\mathrm{H}(t)$, while
$(\Delta\mathrm{\dot{H}})^{2}$ denotes its dispersion. In an alternative
approach presented in Ref. \cite{AC}, an upper bound for the acceleration of
finite dimensional quantum systems in projective Hilbert spaces is derived
utilizing the Robertson uncertainty relation \cite{robertson} for the
evolution of a quantum system under a nonstationary Hamiltonian. A discussion
comparing these two methodologies was presented in Ref. \cite{CB}. The quantum
acceleration threshold states that the modulus of the rate of change in time
of the standard deviation of the time-dependent Hamiltonian operator
$\mathrm{H}=\mathrm{H}\left(  t\right)  $ is upper bounded by the standard
deviation of the time-derivative of the Hamiltonian,
\begin{equation}
\left\vert \frac{d\sigma_{\mathrm{H}}}{dt}\right\vert \leq\sigma
_{\mathrm{\dot{H}}}\text{,} \label{ineq1}%
\end{equation}
with $\mathrm{\dot{H}}\overset{\text{def}}{=}d\mathrm{H}/dt$. Bounds on energy
fluctuations are crucial for comprehending the performance of
quantum-mechanical systems through thermodynamic approaches
\cite{esposito09,garrahan13}. For example, assessing the charging efficiency
of quantum batteries used to introduce, store, and retrieve energy from a
quantum system from a quantum thermodynamic perspective, or evaluating the
cooling power of quantum refrigerators designed to lower a quantum system's
temperature to its minimum, is of paramount significance. This importance is
further underscored by the necessity for high accuracy in quantum
technologies, where accuracy is quantitatively determined by the extent of
fluctuations in any relevant observed quantity, which can be as significant as
their average values at the nanoscale \cite{rinaldi24}. In summary, a
significant level of accuracy requires that fluctuations must not increase excessively.

In this paper, we aim to extend the inequality in Eq. (\ref{ineq1}) to
arbitrary observables $A$ of any finite-dimensional quantum system in a pure
state whose dynamics is governed by a unitary dynamics,
\begin{equation}
\left\vert \frac{d\sigma_{A}}{dt}\right\vert \leq\sigma_{v_{A}}\text{,}
\label{ineq22}%
\end{equation}
where $v_{A}$ denotes a suitably defined velocity observable as we shall see.
For an interesting discussion on the concept of time derivative of a quantum
observable, we suggest Ref. \cite{Fu}. We stress that Eq. (\ref{ineq22})\ is
the main theoretical result obtained by Hamazaki in Ref. \cite{H2}. However,
it is important to emphasize that, unlike Hamazaki's proof, our proposed
derivation is restricted to unitary dynamics, utilizing proof techniques that
illustrate the constraints of quantum acceleration. To date, these constraints
have been examined solely within the context of closed quantum systems.
Nonetheless, our derivation offers a clear elucidation that, at a fundamental
level, the upper limits on the increase of observable fluctuations are
intrinsically linked to the standard uncertainty relations of quantum
mechanics. Ultimately, our detailed study, which incorporates clear
illustrative examples, leads to an inequality that fundamentally assesses the
degree to which quantum mechanics constrains our ability to simultaneously
observe and control both the mean values and fluctuations of observables in
quantum systems that evolve unitarily. This constraint may, in turn, open
avenues for new and significant lines of inquiry within the ever-expanding
field of quantum fluctuations and uncertainty relations in nonequilibrium
thermodynamics \cite{campisi,landi}.

The rest of the paper is organized as follows. In Section II, we revisit
Hamazaki's derivation of the inequality in Eq. (\ref{ineq22}). In Section III,
we derive the inequality given in Eq. (\ref{ineq22}) employing the same
approach used in Refs. \cite{AC,CB}. In Section IV, we discuss our
illustrative examples to verify the inequality. In Section V, we present our
conclusive remarks. Finally, more technical details are located in Appendices A-E.

\section{Hamazaki's Proof Revisited}

In this section, focusing on unitary quantum dynamics in the Schr\"{o}dinger
picture, we present a revisitation of Hamazaki's main theoretical result in
Ref. \cite{H2}. In particular, we focus on quantum unitary dynamics and want
to show that the speed of the standard deviation of an observable $A$ is upper
bounded by the standard deviation of an appropriately defined velocity
observable $v_{A}$. Specifically, we wish to verify the inequality $\left\vert
d\sigma_{A}/dt\right\vert \leq\sigma_{v_{A}}$, where $\sigma_{A}^{2}%
\overset{\text{def}}{=}\left\langle A^{2}\right\rangle -\left\langle
A\right\rangle ^{2}$, $v_{A}$ is such that $\left\langle v_{A}\right\rangle
\overset{\text{def}}{=}d\left\langle A\right\rangle /dt$, and $\left\langle
\cdot\right\rangle $ denotes the expectation value with respect to the quantum
state of interest. We remark that for an explicitly time-dependent observable
$A\left(  t\right)  $, its expectation values in the Schr\"{o}dinger and
Heisenberg pictures are given by $\left\langle A\right\rangle
_{\mathrm{Schr\ddot{o}dinger}}\overset{\text{def}}{=}\left\langle \psi\left(
t\right)  \left\vert A_{\mathrm{S}}\left(  t\right)  \right\vert \psi\left(
t\right)  \right\rangle $ and $\left\langle A\right\rangle
_{\mathrm{Heisenberg}}\overset{\text{def}}{=}\left\langle \psi\left(
0\right)  \left\vert A_{\mathrm{H}}\left(  t\right)  \right\vert \psi\left(
0\right)  \right\rangle $, respectively, where $A_{\mathrm{S}}\left(
t\right)  \overset{\text{def}}{=}A\left(  t\right)  $, $A_{\mathrm{H}}\left(
t\right)  \overset{\text{def}}{=}U^{\dagger}\left(  t\right)  A(t)U(t)$, and
$\left\vert \psi\left(  t\right)  \right\rangle \overset{\text{def}}%
{=}U(t)\left\vert \psi\left(  0\right)  \right\rangle $. For more details, we
suggest consulting Appendix B\textbf{. }Before moving to the proof of Eq.
(\ref{ineq22}), we observe that this inequality can be conveniently rewritten
as%
\begin{equation}
\left(  \frac{d\sigma_{A}}{dt}\right)  ^{2}+\left(  \frac{d\left\langle
A\right\rangle }{dt}\right)  ^{2}\leq\left\langle v_{A}^{2}\right\rangle
\text{.} \label{a2}%
\end{equation}
Before deriving Eq. (\ref{a2}), it is worth observing here that it suggests
the combined squares of the rates of change ($\dot{\mu}_{A}$ and $\dot{\sigma
}_{A}$) of the mean $\mu_{A}\overset{\text{def}}{=}\left\langle A\right\rangle
$ and the standard deviation $\sigma_{A}$ of an observable $A$ are constrained
by the expected value of the square of its corresponding velocity observable
$v_{A}$ (i.e., $\left(  d\mu_{A}/dt\right)  ^{2}+\left(  d\sigma
_{A}/dt\right)  ^{2}\leq\left\langle v_{A}^{2}\right\rangle $). This
constraint signifies that the rates of change for both the mean and the
standard deviation of an observable cannot fluctuate freely, as their squared
sum must remain less than the mean of the square of the velocity observable.
Returning to its derivation, one can arrive at Eq. (\ref{a2}) from Eq.
(\ref{ineq22}) by squaring Eq. (\ref{ineq22}), noting that $\sigma_{v_{A}}%
^{2}=\left\langle v_{A}^{2}\right\rangle -\left\langle v_{A}\right\rangle
^{2}$, and using the definition $\left\langle v_{A}\right\rangle
\overset{\text{def}}{=}d\left\langle A\right\rangle /dt$. In Ref. \cite{H2},
Hamazaki proves Eq. (\ref{ineq22}) by exploiting the equality%
\begin{equation}
\frac{d\left\langle \delta A^{2}\right\rangle }{dt}=2\left\langle \delta
A\text{, }\delta v_{A}\right\rangle \text{,} \label{a3}%
\end{equation}
where $\delta A\overset{\text{def}}{=}A-\left\langle A\right\rangle $ is the
fluctuation observable and\textbf{ }\textrm{cov}$\left(  A\text{, }B\right)
=\left\langle A\text{, }B\right\rangle \overset{\text{def}}{=}\left\langle
\left\{  A\text{, }B\right\}  \right\rangle /2-\left\langle A\right\rangle
\left\langle B\right\rangle $ is the covariance of any two observables $A$ and
$B$. As a side remark, note that \textrm{var}$\left(  A\right)  =\left\langle
\delta A\text{, }\delta A\right\rangle \overset{\text{def}}{=}\left\langle
\delta A^{2}\right\rangle =\sigma_{A}^{2}$, \textrm{var}$\left(  v_{A}\right)
=\left\langle \delta v_{A}\text{, }\delta v_{A}\right\rangle \overset
{\text{def}}{=}\left\langle \delta v_{A}^{2}\right\rangle =\sigma_{v_{A}}^{2}%
$, and $\mathrm{cov}\left(  \delta A\text{, }\delta B\right)  =\left\langle
\delta A\text{, }\delta B\right\rangle \overset{\text{def}}{=}(1/2)\left(
\left\langle \left\{  \delta A\text{, }\delta B\right\}  \right\rangle
\right)  $ since $\left\langle \delta A\right\rangle =\left\langle \delta
B\right\rangle =0$. Postponing for the moment the derivation of Eq.
(\ref{a3}), we point out that one can obtain Eq. (\ref{ineq22}) by using Eq.
(\ref{a3}) together with the following two relations%
\begin{equation}
\left\vert \left\langle \delta A\text{, }\delta v_{A}\right\rangle \right\vert
\leq\sqrt{\left\langle \delta A\text{, }\delta A\right\rangle }\sqrt
{\left\langle \delta v_{A}\text{, }\delta v_{A}\right\rangle }=\sqrt
{\left\langle \delta A^{2}\right\rangle }\sqrt{\left\langle \delta v_{A}%
^{2}\right\rangle }=\sigma_{A}\sigma_{v_{A}}\text{,} \label{a4}%
\end{equation}
and%
\begin{equation}
\frac{d\left\langle \delta A^{2}\right\rangle }{dt}=\frac{d\left(  \sigma
_{A}^{2}\right)  }{dt}=2\sigma_{A}\frac{d\sigma_{A}}{dt}\text{.} \label{a5}%
\end{equation}
Observe that Eq. (\ref{a4}) is just a Cauchy-Schwarz inequality
\cite{steele04}. Finally, simple algebraic manipulations of Eqs. (\ref{a3}),
(\ref{a4}), and (\ref{a5}) yield the inequality in Eq. (\ref{ineq22}).

Let us go back to the proof of Eq. (\ref{ineq22}). Firstly, Hamazaki defines
the expectation value of an observable $A$ as $\left\langle A\right\rangle
\overset{\text{def}}{=}\left(  A\left\vert \rho\right.  \right)  $ with $\rho$
being some probability density and $\left(  \cdot\left\vert \cdot\right.
\right)  $ denoting some inner product. For quantum systems, for instance,
$\rho$ is the density matrix and $\left\langle A\right\rangle \overset
{\text{def}}{=}\mathrm{tr}\left[  A\left(  t\right)  \rho\left(  t\right)
\right]  $. Secondly, Hamazaki assumes that the temporal evolution of
$\rho\left(  t\right)  $ is specified by the relation $d\rho/dt=\mathcal{%
\mathcal{L}%
}\left[  \rho\right]  $ where the map $\mathcal{%
\mathcal{L}%
}\left[  \cdot\right]  $ generally depends on $\rho$ and is not unique. For a
formal discussion on Lindbladian operator $\mathcal{L}$, we suggest Refs.
\cite{Lind,milburn94,scully97}. Thirdly, Hamazaki defines the dual map
$\mathcal{%
\mathcal{L}%
}^{\dagger}$ of $\mathcal{%
\mathcal{L}%
}$ in such a manner that $\left(  A\left\vert \dot{\rho}\right.  \right)
=\left(  A\left\vert \mathcal{%
\mathcal{L}%
}\left[  \rho\right]  \right.  \right)  =\left(  \mathcal{%
\mathcal{L}%
}^{\dagger}\left[  A\right]  \left\vert \rho\right.  \right)  =\left\langle
\mathcal{%
\mathcal{L}%
}^{\dagger}\left[  A\right]  \right\rangle $ for any observable $A$. Finally,
the velocity observable $v_{A}$ is defined as%
\begin{equation}
v_{A}\overset{\text{def}}{=}\dot{A}+\mathcal{%
\mathcal{L}%
}^{\dagger}\left[  A\right]  \text{,} \label{a6}%
\end{equation}
with%
\begin{equation}
\left\langle v_{A}\right\rangle =\left\langle \dot{A}+\mathcal{%
\mathcal{L}%
}^{\dagger}\left[  A\right]  \right\rangle =\left\langle \dot{A}\right\rangle
+\left\langle \mathcal{%
\mathcal{L}%
}^{\dagger}\left[  A\right]  \right\rangle =\left(  \dot{A}\left\vert
\rho\right.  \right)  +\left(  A\left\vert \dot{\rho}\right.  \right)
=\frac{d}{dt}\left(  A\left\vert \rho\right.  \right)  =\frac{d\left\langle
A\right\rangle }{dt}\text{.} \label{aa6}%
\end{equation}
Note that in Eqs. (\ref{a6}), (\ref{aa6}) $\dot{A}$ denotes $\partial
{A}/\partial{t}$. At this point, to derive Eq. (\ref{a3}) given the velocity
observable $v_{A}$ in Eq. (\ref{a6}), we observe that%
\begin{align}
\frac{d\left\langle \delta A^{2}\right\rangle }{dt}  &  =\frac{d}{dt}\left(
\delta A^{2}\left\vert \rho\right.  \right) \nonumber\\
&  =\left(  \left\{  \delta A\text{, }\dot{A}\right\}  \left\vert \rho\right.
\right)  +\left(  A^{2}\left\vert \dot{\rho}\right.  \right)  -2\left\langle
A\right\rangle \left(  A\left\vert \dot{\rho}\right.  \right)  +\left\langle
A\right\rangle ^{2}\left(  \mathbf{1}\left\vert \dot{\rho}\right.  \right)
\nonumber\\
&  =\left(  \left\{  \delta A\text{, }\dot{A}\right\}  \left\vert \rho\right.
\right)  +\left(  \mathcal{%
\mathcal{L}%
}^{\dagger}\left[  A^{2}\right]  \left\vert \rho\right.  \right)
-2\left\langle A\right\rangle \left(  \mathcal{%
\mathcal{L}%
}^{\dagger}\left[  A\right]  \left\vert \rho\right.  \right)  +\left\langle
A\right\rangle ^{2}\left(  \mathcal{%
\mathcal{L}%
}^{\dagger}\left[  \mathbf{1}\right]  \left\vert \rho\right.  \right)
\nonumber\\
&  =\left(  \left\{  \delta A\text{, }\dot{A}\right\}  +\mathcal{%
\mathcal{L}%
}^{\dagger}\left[  A^{2}\right]  -2\left\langle A\right\rangle \mathcal{%
\mathcal{L}%
}^{\dagger}\left[  A\right]  \left\vert \rho\right.  \right)  \text{,}%
\end{align}
that is,%
\begin{equation}
\frac{d\left\langle \delta A^{2}\right\rangle }{dt}=\left(  \left\{  \delta
A\text{, }\dot{A}\right\}  +\mathcal{%
\mathcal{L}%
}^{\dagger}\left[  A^{2}\right]  -2\left\langle A\right\rangle \mathcal{%
\mathcal{L}%
}^{\dagger}\left[  A\right]  \left\vert \rho\right.  \right)  \text{,}
\label{a7}%
\end{equation}
since $\mathcal{%
\mathcal{L}%
}^{\dagger}\left[  \mathbf{1}\right]  $ vanishes with $\mathbf{1}$ being the
identity operator. Indeed, we observe that $\left\langle \mathcal{%
\mathcal{L}%
}^{\dagger}\left[  \mathbf{1}\right]  \right\rangle =\left(  \mathcal{%
\mathcal{L}%
}^{\dagger}\left[  \mathbf{1}\right]  \left\vert \rho\right.  \right)
=\left(  \mathbf{1}\left\vert \mathcal{%
\mathcal{L}%
}\left[  \rho\right]  \right.  \right)  =\left(  \mathbf{1}\left\vert
\dot{\rho}\right.  \right)  =\mathrm{tr}\left[  \mathbf{1}\dot{\rho}\right]
=d\mathrm{tr}\left[  \rho\right]  /dt=d(1)/dt=0$ for any $\rho$. Clearly,
$\left\{  \cdot\text{, }\cdot\right\}  $ in Eq. (\ref{a7}) denotes the quantum
anti-commutator. Notice that in transitioning from the third to the fourth
line of Eq. (\ref{a6}) we used the fact that $\left(  \left(  A-\left\langle
A\right\rangle \right)  \frac{d\left\langle A\right\rangle }{dt}|\rho\right)
=\frac{d\left\langle A\right\rangle }{dt}\left(  \left(  A-\left\langle
A\right\rangle \right)  |\rho\right)  =0$. Therefore, to prove Eq. (\ref{a3})
given Eq. (\ref{a7}), we need to verify that
\begin{equation}
\left\langle \left\{  \delta A\text{, }\dot{A}\right\}  +\mathcal{%
\mathcal{L}%
}^{\dagger}\left[  A^{2}\right]  -2\left\langle A\right\rangle \mathcal{%
\mathcal{L}%
}^{\dagger}\left[  A\right]  \right\rangle =\left\langle \left\{  \delta
A\text{, }\delta v_{A}\right\}  \right\rangle \text{,}%
\end{equation}
since $2\mathrm{cov}\left(  \delta A\text{, }\delta v_{A}\right)
=2\left\langle \delta A\text{, }\delta v_{A}\right\rangle \overset{\text{def}%
}{=}\left\langle \left\{  \delta A\text{, }\delta v_{A}\right\}  \right\rangle
$. Alternatively, since $\left\langle \left\{  \delta A\text{, }\delta
v_{A}\right\}  \right\rangle =\left\langle \left\{  \delta A\text{, }%
v_{A}\right\}  \right\rangle $, we can check the correctness of the following
relation%
\begin{equation}
\left\langle \left\{  \delta A\text{, }\dot{A}\right\}  +\mathcal{%
\mathcal{L}%
}^{\dagger}\left[  A^{2}\right]  -2\left\langle A\right\rangle \mathcal{%
\mathcal{L}%
}^{\dagger}\left[  A\right]  \right\rangle =\left\langle \left\{  \delta
A\text{, }v_{A}\right\}  \right\rangle \text{.} \label{a8}%
\end{equation}
In Ref. \cite{H2}, Hamazaki claims that Eq. (\ref{a8}) is satisfied provided
that $\mathcal{%
\mathcal{L}%
}^{\dagger}\left[  A^{2}\right]  =\left\{  A\text{, }\mathcal{%
\mathcal{L}%
}^{\dagger}\left[  A\right]  \right\}  $. Indeed, assuming the validity of
this latter relation, one notices that
\begin{equation}
\left\langle \left\{  \delta A\text{, }\dot{A}\right\}  +\mathcal{%
\mathcal{L}%
}^{\dagger}\left[  A^{2}\right]  -2\left\langle A\right\rangle \mathcal{%
\mathcal{L}%
}^{\dagger}\left[  A\right]  \right\rangle =\left\langle \left\{  \delta
A\text{, }\dot{A}\right\}  \right\rangle +\left\langle \left\{  A\text{,
}\mathcal{%
\mathcal{L}%
}^{\dagger}\left[  A\right]  \right\}  \right\rangle -2\left\langle
A\right\rangle \left\langle \mathcal{%
\mathcal{L}%
}^{\dagger}\left[  A\right]  \right\rangle \text{,} \label{a9}%
\end{equation}
and, in addition,%
\begin{align}
\left\langle \left\{  \delta A\text{, }v_{A}\right\}  \right\rangle  &
=\left\langle \left(  \delta A\right)  \left(  v_{A}\right)  +\left(
v_{A}\right)  \left(  \delta A\right)  \right\rangle \nonumber\\
&  =\left\langle \left(  \delta A\right)  \left(  \dot{A}+\mathcal{%
\mathcal{L}%
}^{\dagger}\left[  A\right]  \right)  +\left(  \dot{A}+\mathcal{%
\mathcal{L}%
}^{\dagger}\left[  A\right]  \right)  \left(  \delta A\right)  \right\rangle
\nonumber\\
&  =\left\langle \left\{  \delta A\text{, }\dot{A}\right\}  \right\rangle
+\left\langle \left\{  \delta A\text{, }\mathcal{%
\mathcal{L}%
}^{\dagger}\left[  A\right]  \right\}  \right\rangle \nonumber\\
&  =\left\langle \left\{  \delta A\text{, }\dot{A}\right\}  \right\rangle
+\left\langle \left\{  A\text{, }\mathcal{%
\mathcal{L}%
}^{\dagger}\left[  A\right]  \right\}  \right\rangle -2\left\langle
A\right\rangle \left\langle \mathcal{%
\mathcal{L}%
}^{\dagger}\left[  A\right]  \right\rangle \text{.} \label{a10}%
\end{align}
Comparing Eqs. (\ref{a9}) and (\ref{a10}), we conclude that Eq. (\ref{a8}) is
correct. Thus, Eq. (\ref{a3}) is proven and the inequality in Eq.
(\ref{ineq22}) follows. As a final remark, we point out that for unitary
quantum dynamics, $\mathcal{%
\mathcal{L}%
}^{\dagger}\left[  A\right]  =\left(  i/\hslash\right)  \left[  \mathrm{H}%
\text{, }A\right]  $ and the velocity observable reduces to $v_{A}=\dot
{A}+\left(  i/\hslash\right)  \left[  \mathrm{H}\text{, }A\right]  $, with
$\dot{A}$ being here equal to $\partial{A}/\partial{t}$.

To ensure thoroughness, we emphasize that Hamazaki demonstrated that the speed
limit for fluctuation dynamics can only be achieved in particular
circumstances. A common example occurs when the dual map meets the criterion
$\mathcal{%
\mathcal{L}%
}^{\dagger}\left[  A^{2}\right]  =\left\{  A\text{, }\mathcal{%
\mathcal{L}%
}^{\dagger}\left[  A\right]  \right\}  $. The case of unitary quantum dynamics
falls under this category, where\textbf{ }$\mathcal{%
\mathcal{L}%
}^{\dagger}\left[  A\right]  =(i/\hslash)\left[  \mathrm{H}\text{, }A\right]
$. However, when examining open quantum systems characterized by dissipation
and deviations from unitarity, it becomes evident that\textbf{ }$\mathcal{%
\mathcal{L}%
}^{\dagger}\left[  A\right]  $ does not satisfy the condition $\mathcal{%
\mathcal{L}%
}^{\dagger}\left[  A^{2}\right]  =\left\{  A\text{, }\mathcal{%
\mathcal{L}%
}^{\dagger}\left[  A\right]  \right\}  $, making it impossible to directly
establish a speed limit for the fluctuation dynamics of general open quantum
systems where $\mathcal{%
\mathcal{L}%
}^{\dagger}\left[  A\right]  $ can be represented as $(i/\hslash)\left[
\mathrm{H}\text{, }A\right]  +\sum_{j}l_{j}^{\dagger}Al_{j}-(1/2)\left\{
A\text{, }l\right\}  $, with $l_{j}$ being the so-called Lindblad jump
operator. For further insights into the difficulties encountered when
attempting to extend Hamazaki's analysis to open quantum systems, we direct
readers to Ref. \cite{H2}.

Having revisited Hamazaki's original derivation of the fact that the speed of
the standard deviation of any observable is limited by the standard deviation
of its associated velocity-like observable $v_{A}$, we are now ready to
present our alternative derivation for the inequality given by Eq.
(\ref{ineq22}).

\section{Alternative Proof of the Inequality}

To prove the inequality given in Eq. (\ref{ineq22}), we follow our previous
investigations carried out in Refs. \cite{AC,CB}. In these works, we
introduced the notion of a quantum acceleration limit in projective Hilbert
space for any unitary time evolution of finite-dimensional quantum systems in
a pure state, which evolve under arbitrary nonstationary Hamiltonians. In the
following, we will provide a concise overview of this upper limit, as it has
significantly influenced our alternative proof of Hamazaki's inequality.

Remember that the Fubini-Study infinitesimal line element $ds^{2}$ equals
\cite{anandan90}
\begin{equation}
ds^{2}\overset{\text{def}}{=}4\left[  1-\left\vert \left\langle \psi\left(
t\right)  \left\vert \psi\left(  t+dt\right)  \right.  \right\rangle
\right\vert ^{2}\right]  =\frac{4}{\hslash^{2}}\Delta\mathrm{H}\left(
t\right)  ^{2}dt^{2}\text{,}%
\end{equation}
where $\Delta\mathrm{H}\left(  t\right)  ^{2}\overset{\text{def}}{=}%
\langle\psi\left(  t\right)  \left\vert \mathrm{H}\left(  t\right)
^{2}\right\vert \psi\left(  t\right)  \rangle-\left\langle \psi\left(
t\right)  \left\vert \mathrm{H}\left(  t\right)  \right\vert \psi\left(
t\right)  \right\rangle ^{2}$ denotes\textbf{ }the Hamiltonian uncertainty
$\sigma_{\mathrm{H}}^{2}$ of the system, and $i\hslash\partial_{t}\left\vert
\psi\left(  t\right)  \right\rangle =\mathrm{H}\left(  t\right)  \left\vert
\psi\left(  t\right)  \right\rangle $ specifies the time-dependent
Schr\"{o}dinger equation. The overall distance $s=s(t)$ that the system
traverses within the projective Hilbert space is expressed as%
\begin{equation}
s\left(  t\right)  \overset{\text{def}}{=}\frac{2}{\hslash}\int^{t}%
\Delta\mathrm{H}\left(  t^{\prime}\right)  dt^{\prime}\text{.}%
\end{equation}
Consequently, the transportation speed $v_{\mathrm{H}}\left(  t\right)  $ of
the quantum system within the projective Hilbert space is defined as%
\begin{equation}
v_{\mathrm{H}}\left(  t\right)  \overset{\text{def}}{=}\frac{ds\left(
t\right)  }{dt}=\frac{2}{\hslash}\Delta\mathrm{H}\left(  t\right)  \text{.}
\label{dust}%
\end{equation}
For completeness, we point out that if one defines the Fubini-Study
infinitesimal line element in terms of $ds^{2}\overset{\text{def}}{=}\left[
1-\left\vert \left\langle \psi\left(  t\right)  \left\vert \psi\left(
t+dt\right)  \right.  \right\rangle \right\vert ^{2}\right]  =\left[
\Delta\mathrm{H}\left(  t\right)  ^{2}/\hslash^{2}\right]  dt^{2}$
\cite{sam94,uzdin12}, the transportation speed reduces to $v_{\mathrm{H}%
}\left(  t\right)  \overset{\text{def}}{=}\Delta\mathrm{H}\left(  t\right)
/\hslash$. Making use of\textbf{ }$v_{\mathrm{H}}\left(  t\right)  $ as in Eq.
(\ref{dust}), the quantum acceleration\textbf{ }$a_{\mathrm{H}}(t)$\textbf{
}is defined as%
\begin{equation}
a_{\mathrm{H}}(t)\overset{\text{def}}{=}\frac{dv_{\mathrm{H}}\left(  t\right)
}{dt}=\frac{2}{\hslash}\frac{d\left[  \Delta\mathrm{H}\left(  t\right)
\right]  }{dt}\text{.} \label{explain}%
\end{equation}
However, setting $\hslash=1$\textbf{ }and defining $ds_{\mathrm{FS}}%
^{2}\overset{\text{def}}{=}\Delta\mathrm{H}\left(  t\right)  ^{2}dt^{2}$, the
transportation speed in projective Hilbert space becomes $v_{\mathrm{H}%
}\overset{\text{def}}{=}\sigma_{\mathrm{H}}$, while the acceleration of the
quantum evolution reduces to $a_{\mathrm{H}}\overset{\text{def}}{=}%
\partial_{t}\sigma_{\mathrm{H}}$. Then, for any finite-dimensional quantum
system with a dynamics specified by the time-dependent Hamiltonian
\textrm{H}$(t)$, one can verify that the quantum acceleration limit%
\begin{equation}
\left(  a_{\mathrm{H}}\right)  ^{2}\overset{\text{def}}{=}\left(  \partial
_{t}\sigma_{\mathrm{H}}\right)  ^{2}\leq\left(  \sigma_{\partial_{t}%
\mathrm{H}}\right)  ^{2}\text{.} \label{one}%
\end{equation}
In Refs. \cite{AC,CB}, the inequality presented in Eq. (\ref{one}) was
demonstrated to arise from the Robertson uncertainty relation, which is often
regarded as a generalized uncertainty principle within quantum theory, as it
broadens its applicability to variables that may not be strictly conjugate.

In what follows, inspired by the methods employed in Refs. \cite{AC,CB} to
arrive at Eq. (\ref{one}), we present an alternative proof of Hamazaki's
inequality by extending our methodologies to arbitrary quantum observables
(and not just to energy and Hamiltonian operators). We begin by squaring both
sides of Eq. (\ref{ineq22}) to get
\begin{equation}
\left(  \frac{d\sigma_{A}}{dt}\right)  ^{2}\leq{\sigma_{v_{A}}^{2}}\text{.}
\label{eq:16}%
\end{equation}
Eq. (\ref{eq:16}) implies that the magnitude of the speed of the standard
deviation $\sigma_{A}$ of an observable $A$ is less than that of the standard
deviation $\sigma_{v_{A}}$ associated with its corresponding velocity
observable $v_{A}$ (i.e\textbf{.}, $\left\vert d\sigma_{A}/dt\right\vert
\leq\sigma_{v_{A}}$). This inequality indicates that the fluctuation rate of
an observable in a quantum system is constrained; it cannot surpass the
fluctuation of its corresponding velocity observable. For example, the rate of
energy fluctuation in an isolated quantum system is limited to not exceeding
the fluctuation of the speed of the time-varying Hamiltonian that
characterizes the unitary Schr\"{o}dinger evolution of the closed system under
investigation. For a discussion on the unitarity of more peculiar
quantum-mechanical processes, including the black hole evaporation process, we
suggest Ref. \cite{corda25}.

Before going back to the verification, we point out that in this section we
consider the evolution of observables in the Heisenberg picture (for further
details, including the expressions of the expectation value of an observable
in the Schr\"{o}dinger and Heisenberg pictures, we refer to Appendix B).
Returning to the proof, define an operator $\Delta A\overset{\text{def}}%
{=}A-\left\langle A\right\rangle $ such that $\langle\Delta A^{2}%
\rangle={\sigma_{A}^{2}}$ where ${\sigma_{A}^{2}}$ gives the variance of the
time-dependent operator $A$. Note that $\Delta A$ here is the same as $\delta
A\overset{\text{def}}{=}A-\left\langle A\right\rangle $ used in Hamazaki's
derivation. We substitute Hamazaki's \textquotedblleft$\delta$%
\textquotedblright\textbf{\ }utilized to define the fluctuation operator
$\delta A$ with the symbol\textbf{ }\textquotedblleft$\Delta$%
\textquotedblright. This choice is driven by the observation that
\textquotedblleft$\Delta$\textquotedblright\ is commonly used to ultimately
characterize the dispersion of an observable in quantum mechanics. Similarly,
define $\Delta v_{A}\overset{\text{def}}{=}v_{A}-\left\langle v_{A}%
\right\rangle $ such that $\langle\Delta{v_{A}}^{2}\rangle$ $={\sigma_{v_{A}%
}^{2}}$ where ${\sigma_{v_{A}}^{2}}$ denotes the variance of the
time-dependent operator $v_{A}$. The time derivative of $\sigma_{A}$ can be
written as%
\begin{equation}
\frac{d\sigma_{A}}{dt}=\frac{d}{dt}\left(  \sqrt{\left\langle \Delta
A^{2}\right\rangle }\right)  =\frac{\frac{d}{dt}\langle\Delta A^{2}\rangle
}{2\sqrt{\langle\Delta A^{2}\rangle}}\text{.} \label{eq:17}%
\end{equation}
Squaring both sides of Eq. (\ref{eq:17}) yields
\begin{equation}
\left(  \frac{d\sigma_{A}}{dt}\right)  ^{2}=\frac{\left(  \frac{d\langle\Delta
A^{2}\rangle}{dt}\right)  ^{2}}{4\langle\Delta A^{2}\rangle}\text{,}
\label{eq:18}%
\end{equation}
which can be used\textbf{ }to rewrite Eq. (\ref{eq:16}) as
\begin{equation}
\frac{\left(  \frac{d\langle\Delta A^{2}\rangle}{dt}\right)  ^{2}}%
{4\langle\Delta A^{2}\rangle}\leq\sigma_{v_{A}}^{2}\text{.} \label{eq:19}%
\end{equation}
Rearranging Eq. (\ref{eq:19}) leads to%
\begin{equation}
\langle\Delta A^{2}\rangle\langle{\Delta\dot{A}}^{2}\rangle\geq\frac{1}%
{4}\left(  \frac{d\langle\Delta A^{2}\rangle}{dt}\right)  ^{2}\text{.}
\label{eq:20}%
\end{equation}
Using the following relation,
\begin{equation}
\left(  \frac{d\langle\Delta A^{2}\rangle}{dt}\right)  ^{2}=\left(
\langle\Delta\dot{A}\Delta A+\Delta A\Delta\dot{A}\rangle\right)  ^{2}%
=\langle\left\{  \Delta A\text{, }\Delta\dot{A}\right\}  \rangle^{2}\text{,}
\label{eq:21}%
\end{equation}
we can rewrite Eq. (\ref{eq:20}) as
\begin{equation}
\langle\Delta A^{2}\rangle\langle{\Delta\dot{A}}^{2}\rangle\geq\frac{1}%
{4}\langle\left\{  \Delta A\text{, }\Delta\dot{A}\right\}  \rangle^{2}\text{.}
\label{eq:22}%
\end{equation}
If the inequality given by Eq. (\ref{eq:22}) is correct, we can conclude that
the inequality given by Eq. (\ref{eq:16}) is also true. To show the
correctness of Eq. (\ref{eq:22}), use the uncertainty relation derived from
the Cauchy-Schwarz inequality which is given by $\langle\Delta A^{2}%
\rangle\langle\Delta B^{2}\rangle\geq|\left\langle \Delta A\Delta
B\right\rangle |^{2}$ with $B=\dot{A}$. As a consequence, we realize that it
is enough to show that
\begin{equation}
|\langle\left(  \Delta A\right)  \left(  \Delta\dot{A}\right)  \rangle
|^{2}\geq\frac{1}{4}\langle\left\{  \Delta A\text{, }\Delta\dot{A}\right\}
\rangle^{2}\text{.} \label{eq:25}%
\end{equation}
This can be accomplished by noting that
\begin{equation}
4\left\vert \langle\left(  \Delta A\right)  (\Delta\dot{A})\rangle\right\vert
^{2}=\left\vert \langle\left[  \Delta A\text{, }\Delta\dot{A}\right]
\rangle\right\vert ^{2}+\left\vert \langle\left\{  \Delta A\text{, }\Delta
\dot{A}\right\}  \rangle\right\vert ^{2}\text{.} \label{carlino}%
\end{equation}
Obviously, $\left[  \cdot\text{, }\cdot\right]  $\textbf{ }and $\left\{
\cdot\text{, }\cdot\right\}  $ in Eq. (\ref{carlino}) are the quantum
commutator and the quantum anti-commutator, respectively. For completeness, we
point out that in order to obtain Eq. (\ref{carlino}), we used the fact
that\textbf{ }$\Delta A$\textbf{ }and\textbf{ }$\Delta\dot{A}$\textbf{ }are
observables and hence Hermitian operators\textbf{. }We can now prove the
inequality given by Eq. (\ref{a2}) by rewriting Eq. (\ref{eq:16}) using
$\sigma_{v_{A}}^{2}=\langle{\Delta\dot{A}}^{2}\rangle=\langle\dot{A}%
^{2}\rangle-\langle\dot{A}\rangle^{2}$ as $\langle{\Delta\dot{A}}^{2}\rangle$%

\begin{equation}
\left\vert \frac{d\sigma_{A}}{dt}\right\vert ^{2}\leq\langle\dot{A}^{2}%
\rangle-\langle\dot{A}\rangle^{2}\text{,} \label{32}%
\end{equation}
or, alternatively,%
\begin{equation}
\langle\dot{A}\rangle^{2}+\left\vert \frac{d\sigma_{A}}{dt}\right\vert
^{2}\leq\langle\dot{A}^{2}\rangle\text{.} \label{eq:28}%
\end{equation}
Using $v_{A}\overset{\text{def}}{=}dA/dt=\dot{A}$ and $\left\langle
v_{A}\right\rangle =\langle\dot{A}\rangle=d\left\langle A\right\rangle /dt$,
Eq. (\ref{eq:28}) can be written as
\begin{equation}
\left(  \frac{d\langle A\rangle}{dt}\right)  ^{2}+\left(  \frac{d\sigma_{A}%
}{dt}\right)  ^{2}\leq\langle v_{A}^{2}\rangle\text{.} \label{eq:29}%
\end{equation}

Notably, after presenting our formal derivation, we emphasize that the
inequality expressed in Eq. (\ref{32}) can be regarded as a direct outcome of
the Cauchy-Schwarz inequality concerning covariances
\cite{sampson73,he15,liu19,li22,hossjer22,pelessoni25},%
\begin{equation}
\left\vert \mathrm{cov}\left(  A\text{, }B\right)  \right\vert \leq\sigma
_{A}\sigma_{B}\text{,} \label{covineq}%
\end{equation}
where $A$ and $B$ represent arbitrary observables. Specifically, from
$\sigma_{A}\overset{\text{def}}{=}\sqrt{\langle A^{2}\rangle-\langle
A\rangle^{2}}$, it follows that $d\sigma_{A}/dt=\mathrm{cov}(A$, $\dot
{A})/\sigma_{A}$. By applying this relationship and substituting $B=v_{A}%
=\dot{A}$ in Eq. (\ref{covineq}), we derive our inequality in Eq. (\ref{32}).
More broadly, we observe that by setting $A=A^{\left(  n\right)  }%
\overset{\text{def}}{=}d^{n}A/dt^{n}$ and $B=A^{\left(  n+1\right)  }%
\overset{\text{def}}{=}d^{n+1}A/dt^{n+1}$ in Eq. (\ref{32}) and systematically
repeating the aforementioned reasoning based on the covariance inequality, one
can demonstrate that $\left(  d\sigma_{A^{\left(  n\right)  }}/dt\right)
^{2}\leq\sigma_{A^{\left(  n+1\right)  }}^{2}$ for any\textbf{ }$n\geq0$
\cite{AC}. This latter inequality indicates that the magnitude of the rate of
change of the standard deviation of any $n$-th\ time derivative of an
observable $A$ is constrained by the standard deviation of the $\left(
n+1\right)  $-\textrm{th} time derivative of the same observable $A$. It is
evident that our principal inequality in Eq. (\ref{32}) is derived when $n=0$.

Returning to Eq. (\ref{eq:29}),\textbf{ }we stress that it implies that the
combined squares of the rates of change\textbf{ }($\dot{\mu}_{A}$ and
$\dot{\sigma}_{A}$) of the mean\textbf{ }$\mu_{A}\overset{\text{def}}%
{=}\left\langle A\right\rangle $ and the standard deviation $\sigma_{A}$ of an
observable $A$ are limited by the expected value of the square of its
associated velocity observable $v_{A}$ (i.e., $\left(  d\mu_{A}/dt\right)
^{2}+\left(  d\sigma_{A}/dt\right)  ^{2}\leq\left\langle v_{A}^{2}%
\right\rangle $). This relationship indicates that the rates of change of both
the mean and the standard deviation of an observable are not free to vary
without restriction, as their squared sum is bound to be less than the mean of
the square of the velocity observable. For instance, the quantity
$\langle\mathrm{\dot{H}}^{2}\rangle$ constrains how the average value and the
standard deviation of the energy of an isolated quantum system, which evolves
according to a time-dependent Hamiltonian \textrm{H}$\left(  t\right)  $, may
vary over time.

It is also essential to remember that while the state vector progresses over
time in the Schr\"{o}dinger representation, the observable of the quantum
system changes in the Heisenberg picture of quantum mechanics. Both frameworks
are fundamentally equivalent due to the Stone-von Neumann theorem
\cite{rosenberg04}, which asserts the uniqueness of the canonical commutation
relations between position and momentum operators. Consequently, the two
representations can be viewed merely as a change of basis within Hilbert
space. However, the quantum speed limit cannot be applied to the evolution of
the state when describing quantum dynamics within the Heisenberg picture.
Instead, it is necessary to establish the evolution speed of the observable
for a quantum system in this context \cite{MP}. In Appendix B, we explain in
detail the meaning of the concept of velocity observable $v_{A}\overset
{\text{def}}{=}\dot{A}+\left(  i/\hslash\right)  \left[  \mathrm{H}\text{,
}A\right]  $ (with $\dot{A}\overset{\text{def}}{=}\partial A/\partial t$ in
Hamazaki's notation) in unitary quantum dynamics as originally presented by
Hamazaki in Ref. \cite{H2} along with our viewpoint on $v_{A}\overset
{\text{def}}{=}dA/dt$ used in our derivation.

Despite the formal mathematical similarities, our proof is clearly distinct
from a physics perspective. Specifically, while Hamazaki's proof is broad in
its scope and is based on general statistical equalities that are valid in
both classical and quantum physical frameworks with a robust underlying
probabilistic structure, our proof is more narrowly focused and specifically
depends on the algebra of observables in quantum mechanics. For example, among
other aspects, our proof utilizes the fact that the commutator of two
Hermitian operators results in an anti-Hermitian operator. Consequently, its
expectation value is entirely imaginary. This represents a standard procedure
in quantum mechanics when establishing uncertainty relations. For a schematic
overview that emphasizes the similarities and differences between Hamazaki's
derivation and our proof, both of which lead to upper limits on the growth of
fluctuations in observables within the framework of unitary quantum dynamics,
we direct the reader to Table I.

Having examined an alternative proof demonstrating that the speed of the
standard deviation of any observable is constrained by the standard deviation
of its corresponding velocity-like observable $v_{A}$, we are now prepared to
showcase its relevance through straightforward examples related to the unitary
dynamics of both two-level quantum systems and higher-dimensional physical
systems.\begin{table}[t]
\centering
\begin{tabular}
[c]{c|c|c|c|c}\hline\hline
\textbf{Derivations} & \textbf{Fluctuation observable} & \textbf{Velocity
observable} & \textbf{Mathematics} & \textbf{\ Physics}\\\hline
Hamazaki & $\delta A\overset{\text{def}}{=}A-\left\langle A\right\rangle $ &
$v_{A}\overset{\text{def}}{=}\dot{A}+\mathcal{L}^{\dagger}\left[  A\right]  $
& Cauchy-Schwarz inequality & General statistical equalities\\\hline
Cafaro et \textit{al}. & $\Delta A\overset{\text{def}}{=}A-\left\langle
A\right\rangle $ & $v_{A}\overset{\text{def}}{=}\frac{\partial A}{\partial
t}+\frac{i}{\hslash}\left[  \mathrm{H}\text{, }A\right]  $ & Cauchy-Schwarz
inequality & Algebra of quantum observables\\\hline
\end{tabular}
\caption{A schematic overview highlighting the similarities and key
distinctions between Hamazaki's derivation and that of Cafaro et \textit{al}.,
which both lead to upper limits on the growth of fluctuations in observables
within the context of unitary quantum dynamics.}%
\end{table}

\section{Illustrative Examples}

In this section, we provide three illustrative examples of unitary quantum
dynamics where the inequality given by Eq. (\ref{eq:29}) is satisfied. While
the first two examples focus on two-level systems, the third example addresses
higher-dimensional quantum systems. In particular, in the first example, we
discuss a scenario for a two-level quantum system in which the inequality in
Eq. (\ref{eq:29}) reduces to an equality valid at all times during the quantum
evolution. In the second scenario, instead, a strict inequality is generally
valid during the quantum-mechanical evolution of the two-level system. For a
clear presentation of the most general expression of an observable for a qubit
system, we refer to Ref. \cite{S}. Finally, in our third example, we
illustrate the validity of Eq. (\ref{eq:29}) for a harmonic oscillator in a
finite-dimensional Fock space.

\subsection{Two-Level Quantum Systems}

We begin with two-level quantum systems.

\subsubsection{Tight Upper Bound}

In our first example, we begin by considering a two-level quantum system
specified by a time-dependent Hamiltonian \textrm{H}$\left(  t\right)
\overset{\text{def}}{=}\hslash\omega_{0}\cos\left(  \nu_{0}t\right)
\sigma_{z}$, with $\omega_{0}$ and $\nu_{0}$ in $%
\mathbb{R}
_{+}\backslash\left\{  0\right\}  $. This system is a spin -$1/2$ particle or
qubit with sinusoidally modulated energy splitting. Furthermore, using the
Schr\"{o}dinger representation, we assume the explicitly time-dependent
observable to be given by $A\left(  t\right)  \overset{\text{def}}{=}a\left(
t\right)  \sigma_{x}$ with $a\left(  t\right)  \in%
\mathbb{R}
_{+}\backslash\left\{  0\right\}  $. From a physics perspective, an observable
$A$ can be understood as relating to the measurement of the projection of the
spin angular momentum ($\mathbf{s}\overset{\text{def}}{\mathbf{=}}\left(
\hslash/2\right)  \mathbf{\sigma}$, with $\mathbf{\sigma}\overset{\text{def}%
}{\mathbf{=}}\left(  \sigma_{x}\text{, }\sigma_{y}\text{, }\sigma_{z}\right)
$) or the magnetic moment of the electron ($\mathbf{\mu}\overset{\text{def}%
}{\mathbf{=}}\mu_{B}\mathbf{\sigma}$, where $\mu_{B}$ is approximately
$-9.27\times10^{-24}\left[  \mathrm{MKSA}\right]  $, with $\left[
\mathrm{MKSA}\right]  $ representing the International System of Units) along
a specified or variable direction, contingent upon whether the observable is
explicitly time-dependent \cite{fan,han}. More explicitly, the observable
$A\left(  t\right)  \overset{\text{def}}{=}a\left(  t\right)  \sigma_{x}$
provides a time-dependent scaling of the $x$-spin component measurement.
Physically, this models scenarios like spin precession in an alternating
current (AC) magnetic field along the $z$-direction, qubit control under
modulated detuning, or quantum sensing of oscillating signals, where\textbf{
}$\left\langle A\left(  t\right)  \right\rangle $\textbf{ }tracks the
accumulated coherence or signal over time.\textbf{ }We take the initial state
of the system equal to a superposition state $\left\vert \psi\left(  0\right)
\right\rangle =\left\vert +\right\rangle \overset{\text{def}}{=}\left(
\left\vert 0\right\rangle +\left\vert 1\right\rangle \right)  /\sqrt{2}$. A
simple calculation yields the evolved \ state at arbitrary time $t$,%
\begin{equation}
\left\vert \psi\left(  t\right)  \right\rangle =e^{-\frac{i}{\hslash}\int
_{0}^{t}\mathrm{H}\left(  t^{\prime}\right)  dt^{\prime}}\left\vert
\psi\left(  0\right)  \right\rangle =\frac{e^{-i\frac{\omega_{0}}{\nu_{0}}%
\sin\left(  \nu_{0}t\right)  }}{\sqrt{2}}\left\vert 0\right\rangle
+\frac{e^{i\frac{\omega_{0}}{\nu_{0}}\sin\left(  \nu_{0}t\right)  }}{\sqrt{2}%
}\left\vert 1\right\rangle \text{.} \label{fit}%
\end{equation}
From Eq. (\ref{fit}), we obtain that the mean $\left\langle A\right\rangle $
and the standard deviation $\sigma_{A}$ are given by
\begin{equation}
\left\langle A\right\rangle \overset{\text{def}}{=}\left\langle \psi\left(
t\right)  \left\vert A\right\vert \psi\left(  t\right)  \right\rangle
=a\left(  t\right)  \cos\left[  2\frac{\omega_{0}}{\nu_{0}}\sin\left(  \nu
_{0}t\right)  \right]  \text{,} \label{mean}%
\end{equation}
and%
\begin{equation}
\sigma_{A}\overset{\text{def}}{=}\sqrt{\left\langle A^{2}\right\rangle
-\left\langle A\right\rangle ^{2}}=a\left(  t\right)  \sin\left[
2\frac{\omega_{0}}{\nu_{0}}\sin\left(  \nu_{0}t\right)  \right]  \text{,}
\label{sd}%
\end{equation}
respectively. Furthermore, the velocity observable $v_{A}$ reduces to%
\begin{equation}
v_{A}\overset{\text{def}}{=}\frac{\partial A\left(  t\right)  }{\partial
t}+\frac{1}{i\hslash}\left[  A\left(  t\right)  \text{, \textrm{H}}\left(
t\right)  \right]  =\dot{a}(t)\sigma_{x}-2\omega_{0}a\left(  t\right)
\cos\left(  \nu_{0}t\right)  \sigma_{y}\text{.} \label{speed}%
\end{equation}
From Eq. (\ref{speed}), we have that
\begin{equation}
\left\langle v_{A}^{2}\right\rangle =\dot{a}^{2}\left(  t\right)  +4\omega
_{0}^{2}a^{2}\left(  t\right)  \cos^{2}\left(  \nu_{0}t\right)  \text{,}
\label{speed2}%
\end{equation}
since $\left\{  \sigma_{l}\text{, }\sigma_{m}\right\}  =2\delta_{lm}%
\mathbf{1}$. Finally, inserting Eqs. (\ref{mean}), (\ref{sd}), and
(\ref{speed2}) into Eq. (\ref{eq:29}), we have%
\begin{equation}
\left(  \frac{d\left\{  a\left(  t\right)  \cos\left[  2\frac{\omega_{0}}%
{\nu_{0}}\sin\left(  \nu_{0}t\right)  \right]  \right\}  }{dt}\right)
^{2}+\left(  \frac{d\left\{  a\left(  t\right)  \sin\left[  2\frac{\omega_{0}%
}{\nu_{0}}\sin\left(  \nu_{0}t\right)  \right]  \right\}  }{dt}\right)
^{2}=\dot{a}^{2}\left(  t\right)  +4\omega_{0}^{2}a^{2}\left(  t\right)
\cos^{2}\left(  \nu_{0}t\right)  \text{,} \label{equal1}%
\end{equation}
for any instant $t$ and, in addition, for any choice of $a\left(  t\right)  $,
$\omega_{0}$, and $\nu_{0}$. The equality in Eq. (\ref{equal1}) can be checked
analytically. Eq. (\ref{equal1}) indicates that the expected value of the
square of the velocity observable $v_{A}$ is precisely equivalent to the sum
of the squares of the rates of change of the mean $\mu_{A}$ and the standard
deviation $\sigma_{A}$ of the observable $A$ at any moment throughout the
quantum evolution. Consequently, we refer to this as a tight upper bound. For
further investigations into the tightness of quantum speed limits- quantified
either by the energy spread and average energy of a quantum system, or by the
rate of change of its phase- we refer the reader to Refs. \cite{LT} and
\cite{sun21}, respectively.

To ensure thoroughness, we highlight that distinguishing between scenarios
that produce a tight upper bound and those that result in a loose upper bound
seems to be more difficult for general observables compared to the situation
where the observable is the Hamiltonian. Difficulties arise even in the
context of two-level quantum systems. For example, consider the density matrix
of the system at time $t$ represented by $\rho\left(  t\right)  \overset
{\text{def}}{=}(\left[  \mathbf{1+a}\left(  t\right)  \mathbf{\cdot\sigma
}\right]  \mathbf{/}2$, the generally time-dependent Hamiltonian defined by
\textrm{H}$\left(  t\right)  \overset{\text{def}}{=}\mathbf{h}\left(
t\right)  \cdot\mathbf{\sigma}$, and an observable of choice denoted by
$M\left(  t\right)  \overset{\text{def}}{=}\mathbf{m}\left(  t\right)
\cdot\mathbf{\sigma}$. Subsequently, building on our previous work in Ref.
\cite{CB}, one can reformulate the inequalities $\left(  d\sigma
_{A}/dt\right)  ^{2}\leq\sigma_{v_{A}}^{2}$ or, alternatively, $\left(
d\langle M\rangle/dt\right)  ^{2}+\left(  d\sigma_{M}/dt\right)  ^{2}%
\leq\langle v_{M}^{2}\rangle$, in terms of a geometric inequality that
involves dot and cross products with the Bloch vector $\mathbf{a}\left(
t\right)  $,\textbf{ }the magnetic field vector $\mathbf{h}\left(  t\right)
$, and the observable-related vector $\mathbf{m}\left(  t\right)  $. Indeed,
after performing some algebra and setting $\hslash=1$, one can derive
$v_{M}=\left[  \mathbf{\dot{m}+}2\mathbf{m\times h}\right]  \cdot
\mathbf{\sigma}$, $v_{M}^{2}=\left[  \mathbf{\dot{m}+}2\mathbf{m\times
h}\right]  ^{2}\mathbf{1}$, $\langle M\rangle=\mathbf{a\cdot m}$, $\sigma
_{M}^{2}=\mathbf{m}^{2}-\left(  \mathbf{a\cdot m}\right)  ^{2}$, $\left\langle
v_{M}\right\rangle =\mathbf{a\cdot}\left[  \mathbf{\dot{m}+}2\mathbf{m\times
h}\right]  $, and $\left\langle v_{M}^{2}\right\rangle =\left[  \mathbf{\dot
{m}+}2\mathbf{m\times h}\right]  ^{2}$. Thereafter, one can utilize these
geometric relationships to reformulate the aforementioned inequalities and
seek the appropriate geometric configurations among the vectors $\mathbf{a}$,
$\mathbf{h}$, and $\mathbf{m}$ that yield either tight or loose upper bounds
(for more details on the physics of these geometric configurations, see
Appendix C). As stated earlier, this task seems to be intricate when the
observable is not the Hamiltonian. This intricacy emerges due to the increase
in the number of vectors involved, transitioning from two vectors $\left\{
\mathbf{a}\text{, }\mathbf{h}\right\}  $ with $\mathbf{\dot{a}=2h\times a}$ to
a collection of three vectors $\left\{  \mathbf{a}\text{, }\mathbf{h}\text{,
}\mathbf{m}\right\}  $. In this latter scenario, the Bloch vector $\mathbf{a}$
and the observable-related vector $\mathbf{m}$ typically do not conform to a
specified differential equation. In cases where the observable is the
Hamiltonian, suitable configurations have been documented by some of us in
Ref. \cite{CB}. In this paper, however, we have made educated guesses. We
aspire to develop a more systematic approach to differentiate between these
configurations in our future endeavors.

Next, we will consider an example of a loose upper bound.\begin{figure}[t]
\centering
\includegraphics[width=0.50\textwidth] {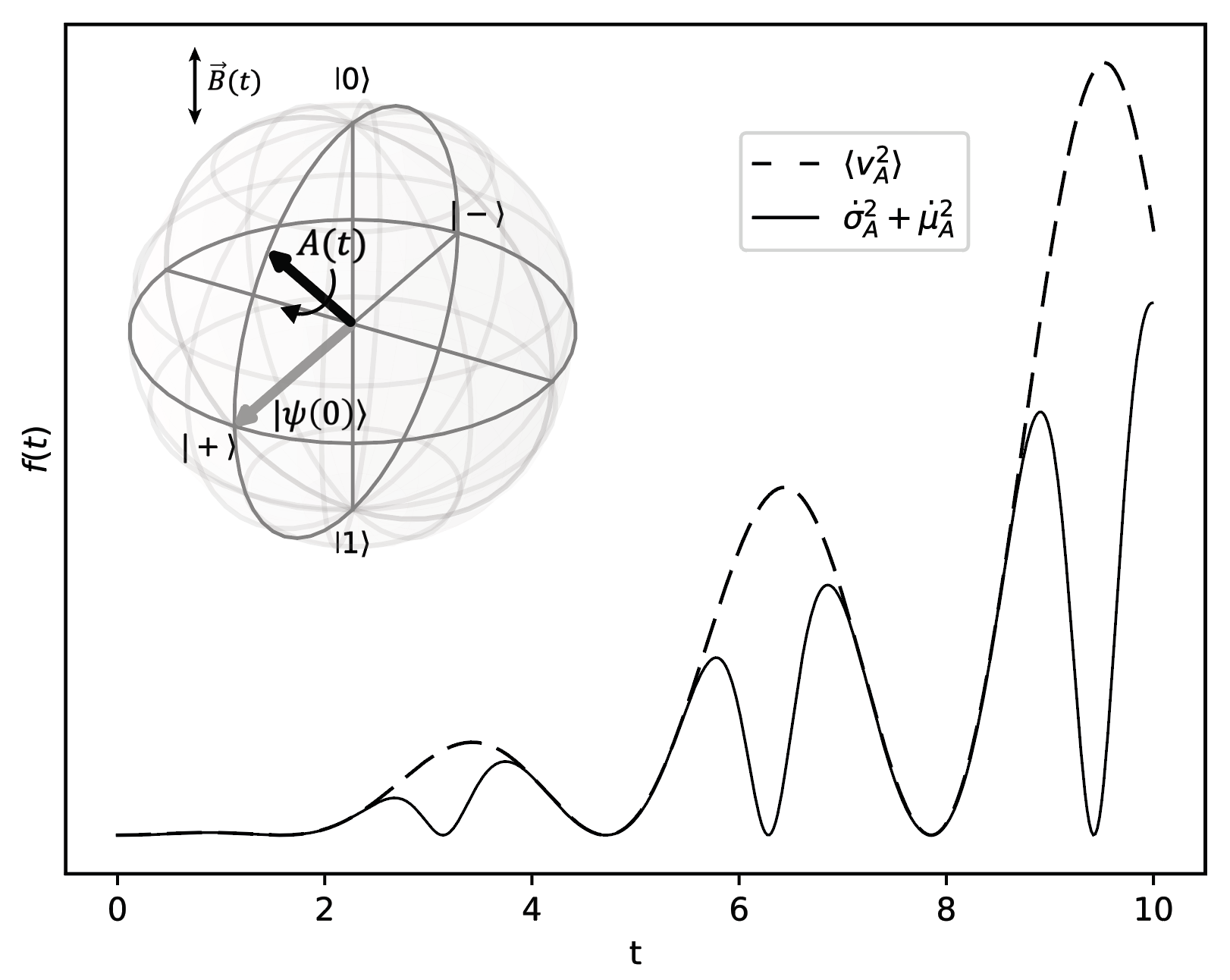}\caption{Numerical verification
of the inequality $(\dot{\sigma}_{A})^{2}+\left(  \dot{\mu}_{A}\right)
^{2}\leq\left\langle v_{A}^{2}\right\rangle $ with $\mu_{A}\overset
{\text{def}}{=}\left\langle A\right\rangle $ for $\omega_{0}=\nu_{0}=1$ and
$A\left(  t\right)  \overset{\text{def}}{=}a\left(  t\right)  \sigma
_{x}+b\left(  t\right)  \sigma_{z}$, with $a\left(  t\right)  =b\left(
t\right)  =t$. Observe that $\left\langle v_{A}^{2}\right\rangle $ and
$(\dot{\sigma}_{A})^{2}+\left(  \dot{\mu}_{A}\right)  ^{2}$ are represented by
a dashed and a solid line, respectively. Formally, the evolution of the
two-level quantum system on the Bloch sphere, starting from the initial state
$\left\vert \psi\left(  0\right)  \right\rangle =\left\vert +\right\rangle
\overset{\text{def}}{=}(\left\vert 0\right\rangle +\left\vert 1\right\rangle
)/\sqrt{2}$, can be viewed as specified by a time-dependent Hamiltonian of the
form \textrm{H}$\left(  t\right)  \overset{\text{def}}{=}-\vec{\mu}\cdot
\vec{B}\left(  t\right)  $, where $\vec{B}\left(  t\right)  $ is the
time-varying magnetic field, $\vec{\mu}\overset{\text{def}}{=}-\mu
_{\mathrm{B}}\vec{\sigma}$ is the magnetic moment of the electron, and
$\mu_{\mathrm{B}}\overset{\text{def}}{=}e\hslash/(2m_{e})\simeq+9.27\times
10^{-24}$ $\left[  \mathrm{MKSA}\right]  $ is the Bohr magneton.}%
\end{figure}

\subsubsection{Loose Upper Bound}

In our second example, we consider a two-level quantum systems whose dynamics
is described by the time-dependent Hamiltonian \textrm{H}$\left(  t\right)
\overset{\text{def}}{=}\hslash\omega_{0}\cos\left(  \nu_{0}t\right)
\sigma_{z}$, with $\omega_{0}$ and $\nu_{0}$ in $%
\mathbb{R}
_{+}\backslash\left\{  0\right\}  $. We note that the Hamiltonian is the same
in both examples. However, using the Schr\"{o}dinger representation, we assume
now an explicitly time-dependent observable to be given by $A\left(  t\right)
\overset{\text{def}}{=}a\left(  t\right)  \sigma_{x}+b\left(  t\right)
\sigma_{z}$ with $a\left(  t\right)  $ and $b(t)$ belonging to $%
\mathbb{R}
_{+}\backslash\left\{  0\right\}  $. We take the initial state of the system
equal to $\left\vert \psi\left(  0\right)  \right\rangle =\left\vert
+\right\rangle \overset{\text{def}}{=}\left(  \left\vert 0\right\rangle
+\left\vert 1\right\rangle \right)  /\sqrt{2}$. As in the previous example,
the evolved state at arbitrary time $t$ is given in Eq. (\ref{fit}). From the
expression of $\left\vert \psi\left(  t\right)  \right\rangle $ in Eq.
(\ref{fit}), we observe that the mean $\left\langle A\right\rangle $ and the
standard deviation $\sigma_{A}$ become
\begin{equation}
\left\langle A\right\rangle \overset{\text{def}}{=}\left\langle \psi\left(
t\right)  \left\vert A\right\vert \psi\left(  t\right)  \right\rangle
=a\left(  t\right)  \cos\left[  2\frac{\omega_{0}}{\nu_{0}}\sin\left(  \nu
_{0}t\right)  \right]  \text{,} \label{mean2}%
\end{equation}
and%
\begin{equation}
\sigma_{A}\overset{\text{def}}{=}\sqrt{\left\langle A^{2}\right\rangle
-\left\langle A\right\rangle ^{2}}=\sqrt{a^{2}\left(  t\right)  \sin
^{2}\left[  2\frac{\omega_{0}}{\nu_{0}}\sin\left(  \nu_{0}t\right)  \right]
+b^{2}(t)}\text{,} \label{sd2}%
\end{equation}
respectively. In deriving Eq. (\ref{sd2}), we used the anti-commutation rule
for Pauli operators \cite{djo12}, $\left\{  \sigma_{l}\text{, }\sigma
_{m}\right\}  =2\delta_{lm}\mathbf{1}$ with $\mathbf{1}$ being the identity
operator. Moreover, the velocity observable $v_{A}$ is given by
\begin{equation}
v_{A}\overset{\text{def}}{=}\frac{\partial A\left(  t\right)  }{\partial
t}+\frac{1}{i\hslash}\left[  A\left(  t\right)  \text{, \textrm{H}}\left(
t\right)  \right]  =\dot{a}(t)\sigma_{x}+\dot{b}\left(  t\right)  \sigma
_{z}-2\omega_{0}a\left(  t\right)  \cos\left(  \nu_{0}t\right)  \sigma
_{y}\text{.} \label{speed22}%
\end{equation}
Employing Eq. (\ref{speed22}) along with the anti-commutation rule $\left\{
\sigma_{l}\text{, }\sigma_{m}\right\}  =2\delta_{lm}\mathbf{1}$, we obtain
\begin{equation}
\left\langle v_{A}^{2}\right\rangle =\dot{a}^{2}\left(  t\right)  +\dot{b}%
^{2}(t)+4\omega_{0}^{2}a^{2}\left(  t\right)  \cos^{2}\left(  \nu_{0}t\right)
\text{,} \label{speed222}%
\end{equation}
since (as previously mentioned) $\left\{  \sigma_{l}\text{, }\sigma
_{m}\right\}  =2\delta_{lm}\mathbf{1}$. Finally, inserting Eqs. (\ref{mean2}),
(\ref{sd2}), and (\ref{speed222}) into Eq. (\ref{eq:29}), we generally have%
\begin{equation}
\left(  \frac{d\left\{  a\left(  t\right)  \cos\left[  2\frac{\omega_{0}}%
{\nu_{0}}\sin\left(  \nu_{0}t\right)  \right]  \right\}  }{dt}\right)
^{2}+\left(  \frac{d\left\{  \sqrt{a^{2}\left(  t\right)  \sin^{2}\left[
2\frac{\omega_{0}}{\nu_{0}}\sin\left(  \nu_{0}t\right)  \right]  +b^{2}%
(t)}\right\}  }{dt}\right)  ^{2}\leq\dot{a}^{2}\left(  t\right)  +\dot{b}%
^{2}(t)+4\omega_{0}^{2}a^{2}\left(  t\right)  \cos^{2}\left(  \nu_{0}t\right)
\text{,} \label{u}%
\end{equation}
for any choice of $a\left(  t\right)  $, $b(t)$, $\omega_{0}$, and $\nu_{0}$.
Although we are unable of proving analytically the inequality in Eq. (\ref{u})
in the general case, we emphasize that it can be explicitly checked in special
cases. For instance, set $\xi\overset{\text{def}}{=}2(\omega_{0}/\nu_{0}%
)\sin\left(  \nu_{0}t\right)  $. Then, it is straightforward to verify that by
expanding out the derivatives under the square, using\textbf{ }$d\left[
\cos\left(  \xi\right)  \right]  =dt=-\sin\left(  \xi\right)  d\xi/dt$ and
$d\left[  \sin^{2}(\xi)\right]  /dt=2\sin\left(  \xi\right)  \cos\left(
\xi\right)  d\xi/dt$ and, finally, letting $\xi=n\pi$ with $n\in%
\mathbb{Z}
$, we have that $\sin\left(  \xi\right)  \rightarrow0$ and $\cos^{2}\left(
\xi\right)  =1$. Therefore, we obtain for the LHS of Eq. (\ref{u}) $\dot
{a}^{2}+\dot{b}^{2}$, while the RHS is unchanged and equals $\dot{a}^{2}%
+\dot{b}^{2}+4\omega_{0}^{2}a^{2}\cos^{2}\left(  \nu_{0}t\right)  $. For this
reason, the inequality in Eq. (\ref{u}) is clearly satisfied, being\textbf{
}$4\omega_{0}^{2}a^{2}\cos^{2}\left(  \nu_{0}t\right)  \geq0$. Alternatively,
the inequality can be checked numerically for a given choice of the quantities
$a\left(  t\right)  $, $b(t)$, $\omega_{0}$, and $\nu_{0}$, as displayed in
Fig. $1$. It can be observed from this figure that the expected value of the
square of the velocity observable $v_{A}$ does not precisely correspond to the
sum of the squares of the mean $\mu_{A}$ and the standard deviation
$\sigma_{A}$ of the observable $A$ at any given moment during the quantum
evolution. As a result, we designate this as a loose upper bound. It would be
valuable to comprehend the physical significance of the expression
$\left\langle v_{A}^{2}\right\rangle -(\dot{\mu}_{A}^{2}+\dot{\sigma}_{A}%
^{2})$, particularly in terms of predicting which combinations of Hamiltonian
and observable result in a stringent bound, as opposed to those that produce a
more relaxed one.\begin{figure}[t]
\centering
\includegraphics[width=0.75\textwidth] {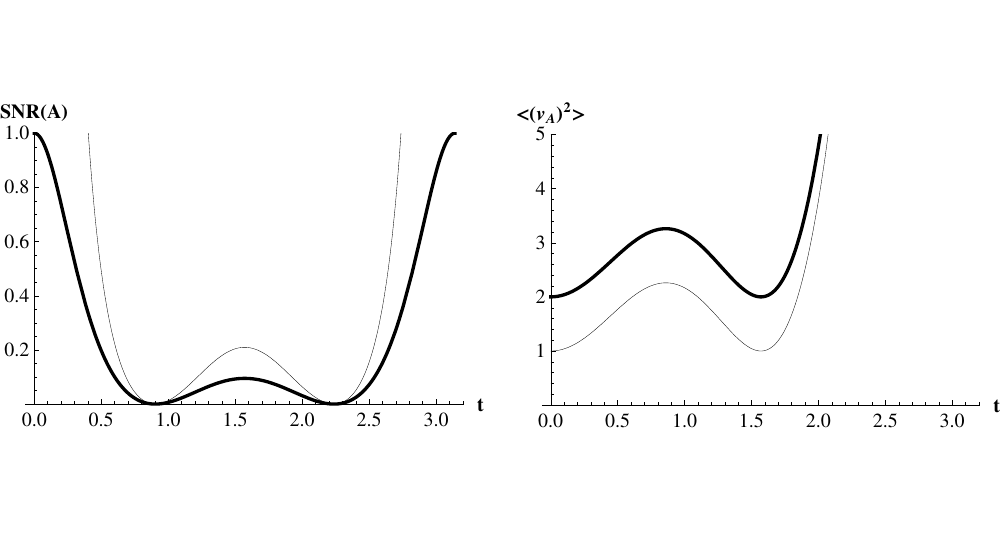}\caption{On the left side, we
plot the \textrm{SNR}$\left(  A\right)  \overset{\text{def}}{=}\left\langle
A\right\rangle ^{2}/\mathrm{var}(A)$ versus time $t$ for the first (thin solid
line) and the second (thick solid line) examples, respectively. On the right
side, instead, we display the temporal behavior of the expectation value of
the square of the velocity observable, $\left\langle v_{A}^{2}\right\rangle
\overset{\text{def}}{=}\left\langle \left(  dA/dt\right)  ^{2}\right\rangle
\geq\left(  \dot{\mu}_{A}\right)  ^{2}+\left(  \dot{\sigma}_{A}\right)  ^{2}$,
for the first (thin solid line) and the second (thick solid line) examples,
respectively. In all plots, we assume $\omega_{0}=\nu_{0}=1$ and $a\left(
t\right)  =b\left(  t\right)  =t$. Recall that the observables $A(t)$ being
measured in the first and second examples are given by $a\left(  t\right)
\sigma_{x}$ and $a\left(  t\right)  \sigma_{x}+b\left(  t\right)  \sigma_{z}$,
respectively. Finally, we point out that the displayed behaviors suggest that
to a lower \textrm{SNR}$\left(  A\right)  $ there corresponds a higher
$\left\langle v_{A}^{2}\right\rangle $.}%
\end{figure}

In conclusion, we highlight that the quality of a signal $S$ can be assessed
through its signal-to-noise ratio (\textrm{SNR}), which is defined as
$\mathrm{S}$\textrm{NR}$\left(  S\right)  \overset{\text{def}}{=}\left\langle
S\right\rangle ^{2}/\mathrm{var}\left(  S\right)  =\mu_{S}^{2}/\sigma_{S}^{2}%
$. The signal $S$ may be characterized, for example, by a stochastic
observable that is associated with the measurement results of its
corresponding (state-dependent) observable at various times. In our
examination of the first and second examples, if we define a similar
\textrm{SNR} for the observables $A$ as \textrm{SNR}$\left(  A\right)
\overset{\text{def}}{=}\left\langle A\right\rangle ^{2}/\mathrm{var}\left(
A\right)  $, we note that%
\begin{equation}
0\leq\frac{\left[  \mathrm{SNR}\left(  A\right)  \right]  _{\mathrm{example}%
\text{-}2}}{\left[  \mathrm{SNR}\left(  A\right)  \right]  _{\mathrm{example}%
\text{-}1}}=\frac{\left[  \mathrm{var}\left(  A\right)  \right]
_{\mathrm{example}\text{-}1}}{\left[  \mathrm{var}\left(  A\right)  \right]
_{\mathrm{example}\text{-}2}}\leq1\text{,} \label{smile}%
\end{equation}
where the equality in Eq. (\ref{smile}) holds true since $\left(  \mu
_{A}\right)  _{\mathrm{example}\text{-}1}=\left(  \mu_{A}\right)
_{\mathrm{example}\text{-}2}$\textbf{. }In our two examinations, we also
have\textbf{ }%
\begin{equation}
0\leq\frac{_{\left\langle v_{A}^{2}\right\rangle _{\mathrm{example}\text{-}1}%
}}{_{\left\langle v_{A}^{2}\right\rangle _{\mathrm{example}\text{-}2}}}%
\leq1\text{.} \label{smile22}%
\end{equation}
Obviously, the denominators in Eqs. (\ref{smile}) and (\ref{smile22}) are
assumed to be nonzero. In the two selected unitary dynamical scenarios
examined here, it is observed that the quality of the signal, as measured by
\textrm{SNR}$\left(  A\right)  $, is enhanced when the overall upper limit on
the sum of the squares of the velocities of $\mu_{A}$ and $\sigma_{A}$,
denoted by $\left\langle v_{A}^{2}\right\rangle $, is reduced. The
inequalities in Eqs. (\ref{smile}) and (\ref{smile22}) are clearly illustrated
in Fig. $2$. To ensure clarity, we highlight that comparing our two examples
is justified, as both involve the examination of the identical unitary
dynamics of the same two-level quantum systems evolving from the same initial
state. The only distinction between the two examples lies in the selection of
the observables we opt to measure. Consequently, we deduce that measuring the
observable $A\left(  t\right)  \overset{\text{def}}{=}a\left(  t\right)
\sigma_{x}$ is anticipated to be less complex in terms of achieving high
accuracy compared to measuring the observable $A\left(  t\right)
\overset{\text{def}}{=}a\left(  t\right)  \sigma_{x}+b\left(  t\right)
\sigma_{z}$ . This conclusion appears to be physically justifiable, as the
first observable can be interpreted as the projection of an electron's
magnetic moment along a fixed direction (specifically, the direction indicated
by the unit vector\textbf{ }$\hat{x}$)\textbf{.} In contrast, the second
observable can be understood as the projection of the magnetic moment of an
electron along a direction that varies with time (particularly, the direction
defined by the unit vector $\left[  a\left(  t\right)  \hat{x}+b\left(
t\right)  \hat{z}\right]  /\sqrt{a^{2}(t)+b^{2}(t)}$). However, additional
observations deserve attention. Firstly, as illustrated in Fig. $2$, when we
compare the numerical estimates of $\left\langle v_{A}^{2}\right\rangle $ for
$A\overset{\text{def}}{=}t\sigma_{x}$ and $A\overset{\text{def}}{=}t\sigma
_{x}+t\sigma_{z}$,\textbf{ }the unit vectors $\hat{n}\left(  t\right)  $ that
appear in $A\overset{\text{def}}{=}\vec{n}\left(  t\right)  \cdot
\mathbf{\sigma}$ with $\vec{n}\left(  t\right)  \overset{\text{def}}%
{=}n(t)\hat{n}\left(  t\right)  $ are represented as $\hat{n}\left(  t\right)
\overset{\text{def}}{=}\hat{x}$\textbf{ }and $\hat{n}\left(  t\right)
\overset{\text{def}}{=}\left(  \hat{x}+\hat{z}\right)  /\sqrt{2}$,
respectively. Consequently, the observable $A\overset{\text{def}}{=}%
t\sigma_{x}+t\sigma_{z}$ is specified by an $\hat{n}\left(  t\right)  $ that
remains constant over time. Nevertheless, we have confirmed that our findings
from Fig\textbf{.} $2$ do not qualitatively alter when comparing $A$
$\overset{\text{def}}{=}t\sigma_{x}$ and $A\overset{\text{def}}{=}t\sigma
_{x}+t^{2}\sigma_{z}$, where the latter is characterized by a time-dependent
unit vector $\hat{n}\left(  t\right)  \overset{\text{def}}{=}(t\hat{x}%
+t^{2}\hat{z})/\sqrt{t^{2}+t^{4}}$. Our analysis indicates that this specific
case reinforces our conclusion that a higher $\left\langle v_{A}%
^{2}\right\rangle $\textbf{ }is associated with a lower \textrm{SNR}$\left(
A\right)  $. Secondly, while the observables chosen in Fig. $2$, specifically
$A\overset{\text{def}}{=}t\sigma_{x}$ and $A\overset{\text{def}}{=}t\sigma
_{x}+t\sigma_{z}$, may formally yield an unphysical asymptotically divergent
expression for $\left\langle v_{A}^{2}\right\rangle $, they exhibit convergent
behavior over finite time intervals relevant to the evolutions under
consideration. Furthermore, we have established that the conclusions presented
in Fig\textbf{.} $2$ hold true even in more physically realistic scenarios
where $\left\langle v_{A}^{2}\right\rangle $ remains convergent at all times.
This situation arises, for example, when conducting a comparative analysis
between the observables\textbf{ }$A\overset{\text{def}}{=}\cos(t)\sigma_{x}$
and $A\overset{\text{def}}{=}\cos(t)\sigma_{x}+\sin(t)\sigma_{z}$ (with
$n(t)=1$\textbf{ }and $\hat{n}\left(  t\right)  =\cos(t)\hat{x}+\sin(t)\hat
{z}$\textbf{ }for the latter observable).

In summary, based on Eqs. (\ref{smile}) and (\ref{smile22}), it seems that
higher fluctuation rates are associated with lower relative qualities of the
signals (for more details, see Appendix D). This observation appears to be
reasonable. Nevertheless, a more comprehensive understanding of this (quantum)
phenomenon necessitates a thorough quantitative analysis, which we will
reserve for future scientific investigations that should also include
thermodynamical arguments on fluctuations
\cite{landauer98,chen14,lutz20,dechiara22,cai24}.

We are now ready to discuss our quantum harmonic oscillator example.

\subsection{Multi-level quantum systems}

In this example, we transition from a two-state system, such as a two-level
atom, to a continuous variables quantum system in an infinite-dimensional
Hilbert space. Specifically, we consider a one-dimensional quantum harmonic
oscillator whose Hamiltonian is defined as%
\begin{equation}
\mathrm{\hat{H}}\overset{\text{def}}{=}\hslash\omega\left(  \hat{a}^{\dagger
}\hat{a}+\frac{1}{2}\right)  \text{.} \label{walid1}%
\end{equation}
In this section, we use the hat-symbol as in Eq. (\ref{walid1}) to denote
operators. In Eq. (\ref{walid1}), $\omega$ is the angular frequency of the
oscillator, while $\hat{a}$ and $\hat{a}^{\dagger}$ are the annihilation and
creation operators, respectively. They are given by,%
\begin{equation}
\hat{a}\overset{\text{def}}{=}\sqrt{\frac{m\omega}{2\hslash}}(\hat{x}%
+i\frac{\hat{p}}{m\omega})\text{, and }\hat{a}^{\dagger}\overset{\text{def}%
}{=}\sqrt{\frac{m\omega}{2\hslash}}(\hat{x}-i\frac{\hat{p}}{m\omega})\text{.}
\label{walid2}%
\end{equation}
From the two relations in Eq. (\ref{walid2}), one can obtain inverse relations
to express the position and the momentum operators $\hat{x}$ and $\hat{p}$,
respectively, as
\begin{equation}
\hat{x}\overset{\text{def}}{=}\sqrt{\frac{\hslash}{2m\omega}}\left(  \hat
{a}+\hat{a}^{\dagger}\right)  \text{, and }\hat{p}\overset{\text{def}}%
{=}i\sqrt{\frac{m\omega\hslash}{2}}\left(  \hat{a}^{\dagger}-\hat{a}\right)
\text{.} \label{walid3}%
\end{equation}
We assume to study the quantum evolution under the Hamiltonian $\mathrm{\hat
{H}}$ in Eq. (\ref{walid1}) of an initial state specified by a displaced
squeezed vacuum state $\left\vert \Psi\left(  0\right)  \right\rangle $ given
by%
\begin{equation}
\left\vert \Psi\left(  0\right)  \right\rangle \overset{\text{def}}{=}\hat
{D}\left(  \alpha\right)  \hat{S}\left(  z\right)  \left\vert 0\right\rangle
\text{,} \label{walid4}%
\end{equation}
where $\hat{D}\left(  \alpha\right)  \overset{\text{def}}{=}e^{\alpha\hat
{a}^{\dagger}-\alpha^{\ast}\hat{a}}$ is the unitary displacement operator,
$\hat{S}\left(  z\right)  \overset{\text{def}}{=}e^{\frac{z^{\ast}}{2}\hat
{a}^{2}-\frac{z}{2}\left(  \hat{a}^{\dagger}\right)  ^{2}}$ is the unitary
squeeze operator, and $\left\vert 0\right\rangle $ is the vacuum state
\cite{gerry05,akira15,agarwal13}. Setting $\hat{D}\left(  \alpha\right)
\hat{S}\left(  z\right)  \left\vert 0\right\rangle \overset{\text{def}}%
{=}\left\vert z\text{, }\alpha\right\rangle $, we stress that the $\left\vert
\Psi\left(  0\right)  \right\rangle =\left\vert z\text{, }\alpha\right\rangle
=\sum_{n=0}^{\infty}c_{n}\left\vert n\right\rangle $\textbf{\ }with
$c_{n}\overset{\text{def}}{=}\left\langle n\left\vert z\text{, }\alpha\right.
\right\rangle $. For an explicit expression of the coefficients $\left\{
c_{n}\right\}  $ that describe the quantum overlap between the number states
$\left\{  \left\vert n\right\rangle \right\}  $ and the squeezed coherent
states $\left\{  \left\vert z\text{, }\alpha\right\rangle \right\}  $ in terms
of Hermite polynomials with complex arguments, we suggest Ref.
\cite{agarwal13}\textbf{. }The displacement operator creates coherent states
by displacing the ground state. The squeeze operator, instead, generates
squeezed states by manipulating the fluctuations of the quadrature fields used
to express optical fields. Observe that $\alpha$ denotes the complex
displacement parameter that specifies the amount of displacement in optical
phase space, while $z$ is an arbitrary complex number with $\left\vert
z\right\vert $ specifying the degree of squeezing. From Eqs. (\ref{walid1})
and (\ref{walid4}), the evolved state at time $t$ is given by $\left\vert
\Psi\left(  t\right)  \right\rangle \overset{\text{def}}{=}e^{-\frac
{i}{\hslash}\mathrm{\hat{H}}t}\left\vert \Psi\left(  0\right)  \right\rangle
$. This is the state that we use to numerically evaluate the expectation
values required to evaluate our inequality $\left(  \dot{\mu}_{A}\right)
^{2}+\left(  \dot{\sigma}_{A}\right)  ^{2}\leq\left\langle v_{A}%
^{2}\right\rangle $. Most importantly, the time-dependent observable that we
choose to consider is defined by $\hat{A}\left(  t\right)  \overset
{\text{def}}{=}\hat{A}\left[  \theta\left(  t\right)  \right]  =\cos\left[
\theta\left(  t\right)  \right]  \hat{x}+\sin\left[  \theta\left(  t\right)
\right]  \hat{p}$, with $\theta\left(  t\right)  \overset{\text{def}}{=}%
\cos(t)$. In summary, the choice of the initial state $\left\vert \Psi\left(
0\right)  \right\rangle \overset{\text{def}}{=}\hat{D}\left(  \alpha\right)
\hat{S}\left(  z\right)  \left\vert 0\right\rangle $\textbf{\ }is motivated by
several key physical and mathematical considerations. Firstly, this displaced
squeezed vacuum state represents a fully general pure Gaussian state in
quantum optics, capable of exhibiting both non-zero displacement (mean field)
and reduced quantum fluctuations in a selected quadrature. Simultaneously it
exhibits non-zero expectation values ($\left\langle \hat{x}\right\rangle
\neq0$, $\left\langle \hat{p}\right\rangle \neq0$)\textbf{\ }through the
displacement $\alpha$, and tunable quantum fluctuations ($\Delta\hat{x}$,
$\Delta\hat{p}$) through the squeezing parameter $z$, allowing us to probe
both aspects of the uncertainty relation. \ Secondly, the state's time
evolution under \textrm{\^{H}} in Eq. (\ref{walid1}) generates non-trivial
dynamics where both the mean values and variances of the quadratures evolve in
non- commensurate ways, providing a rich testbed for the inequality $(\dot
{\mu}_{A})^{2}+\left(  \dot{\sigma}_{A}\right)  ^{2}\leq\left\langle v_{A}%
^{2}\right\rangle $. Lastly, it corresponds to experimentally realizable
states in quantum optics via displacement and squeezing operations, with the
chosen observable $\hat{A}\left(  t\right)  $ directly measurable through
balanced homodyne detection. The time-dependent observable\textbf{ }$\hat
{A}\left(  t\right)  $\textbf{ }mirrors actual experimental configurations
where the measured quadrature rotates in phase space, making our theoretical
analysis directly relevant to quantum optical implementations. The state's
combination of classical displacement and quantum squeezing ensures
non-trivial evolution of both the mean signal and quantum noise
characteristics- precisely the quantities constrained by our fundamental
inequality in Eq. (\ref{eq:29}). Concerning this last point, it is noteworthy
that the selection of\textbf{ }$\hat{A}\left(  t\right)  $ is driven by the
observation that the instantaneous value of $\cos(\theta)\hat{x}+\sin\left(
\theta\right)  \hat{p}$, where $\alpha\overset{\text{def}}{=}\left\vert
\alpha\right\vert e^{i\theta}$ denotes the complex amplitude of the optical
electric field, corresponds to the instantaneous output of a balanced homodyne
measurement in a standard quantum optics experimental configuration
\cite{gerry05,akira15,agarwal13}. The balanced homodyne measurement is
designed to identify squeezed light. The fundamental concept involves
combining the signal field, which is expected to exhibit squeezing, with a
strong coherent field known as the local oscillator, using a $50$:$50$ beam
splitter. The result of this measurement is represented by the difference in
photocurrent between two detectors\textbf{,} $\hat{I}_{2}-\hat{I}_{1}$, where
$\hat{I}_{1}$ and $\hat{I}_{2}$ denote the photon counts in modes $1$ and $2$,
respectively. Moreover, in the field of quantum optics, the variables $\hat
{x}$ and $\hat{p}$ are referred to as quadrature operators, or equivalently,
as generalized position and momentum. While they are conjugate variables
characterized by the relation $\left[  \hat{x}\text{, }\hat{p}\right]
=i\hslash$, it is important to distinguish them from the standard position and
momentum operators used in quantum mechanics. For more details on the balanced
homodyne measurement, we suggest Refs. \cite{gerry05,akira15,agarwal13}.

In what follows, we check the validity of our inequality $(\dot{\mu}_{A}%
)^{2}+\left(  \dot{\sigma}_{A}\right)  ^{2}\leq\left\langle v_{A}%
^{2}\right\rangle $ via an approximate numerical analysis that relies on the
concept of truncated coherent states \cite{mainardi23}. These states are
formulated by treating the Fock space of the quantum harmonic oscillator as
finite-dimensional. This is achieved by limiting the series that represents
coherent states within the infinite-dimensional Fock space. It is important to
note that when the dimensionality of the state space significantly exceeds the
mean occupation number of the coherent states, the findings derived from the
finite-dimensional framework remain valid for a conventional
quantum-mechanical harmonic oscillator. In essence, when the intensity of the
coherent state is considerably less than the dimension of the state space,
both the standard coherent state and the coherent state defined in the
finite-dimensional context exhibit identical statistical and phase
characteristics \cite{buzek92,kuang94}.\textbf{ }More explicitly, in the limit
of $s\gg\left\vert \alpha\right\vert ^{2}$, the mean excitation number\textbf{
}$\left\langle \hat{N}\right\rangle $ approaches $\left\vert \alpha\right\vert
^{2}$, with $\hat{N}\overset{\text{def}}{=}\hat{a}^{\dagger}\hat{a}=\sum
_{n=1}^{s}n\left\vert n\right\rangle \left\langle n\right\vert $ being the
number operator. In other words, $\left\langle \hat{N}\right\rangle
\overset{\left\vert \alpha\right\vert ^{2}\ll s}{\approx}$ $\left\vert
\alpha\right\vert ^{2}$, with $s+1$ being the dimension of the
finite-dimensional (Hilbert) space spanned by the number states\textbf{
}$\left\{  \left\vert 0\right\rangle \text{,..., }\left\vert s\right\rangle
\right\}  $.

Keeping these theoretical remarks in mind, we used the QuTiP Python package
(i.e., an open-source software for simulating the dynamics of quantum systems)
to numerically solve the Schr\"{o}dinger equation of interest\textbf{.} The
solver requires the initial state $\left\vert \Psi\left(  0\right)
\right\rangle $ in Eq. (\ref{walid4}) and the Hamiltonian \textrm{\^{H}} in
Eq. (\ref{walid1}) as inputs. Subsequently, it calculates the system's state
at each designated time step, denoted as $\left\vert \Psi\left(  t\right)
\right\rangle $. From the wavefunction at each time step, one can derive any
desired quantity, including the expectation values of observables. A
significant challenge associated with the quantum harmonic oscillator is that,
theoretically, the Hamiltonian possesses an infinite number of eigenstates,
each corresponding to a specific photon number or a Fock state $\left\vert
n\right\rangle $, where $n\in%
\mathbb{N}
$. However, in numerical simulations, it is necessary to truncate the Hilbert
space by defining a maximum photon number. In this instance, we have
constrained the Hilbert space to $s=20$ photons. This decision is justified,
as the intensity of the state (i.e., $\left\vert \alpha\right\vert ^{2}$)
employed in the simulation is $5$, indicating that, on average, there are
roughly five photons present. Consequently, the probability amplitude for
higher photon numbers remains very small in the $\left(  s+1\right)
$-dimensional Hilbert space. Additionally, to confirm the accuracy of this
truncation, we can assess the normalization of the state at each time step,
ensuring it stays near one. In this instance, the normalization is effectively
preserved, thereby validating the truncation's accuracy (for further details,
see Appendix E).\begin{figure}[t]
\centering
\includegraphics[width=0.75\textwidth] {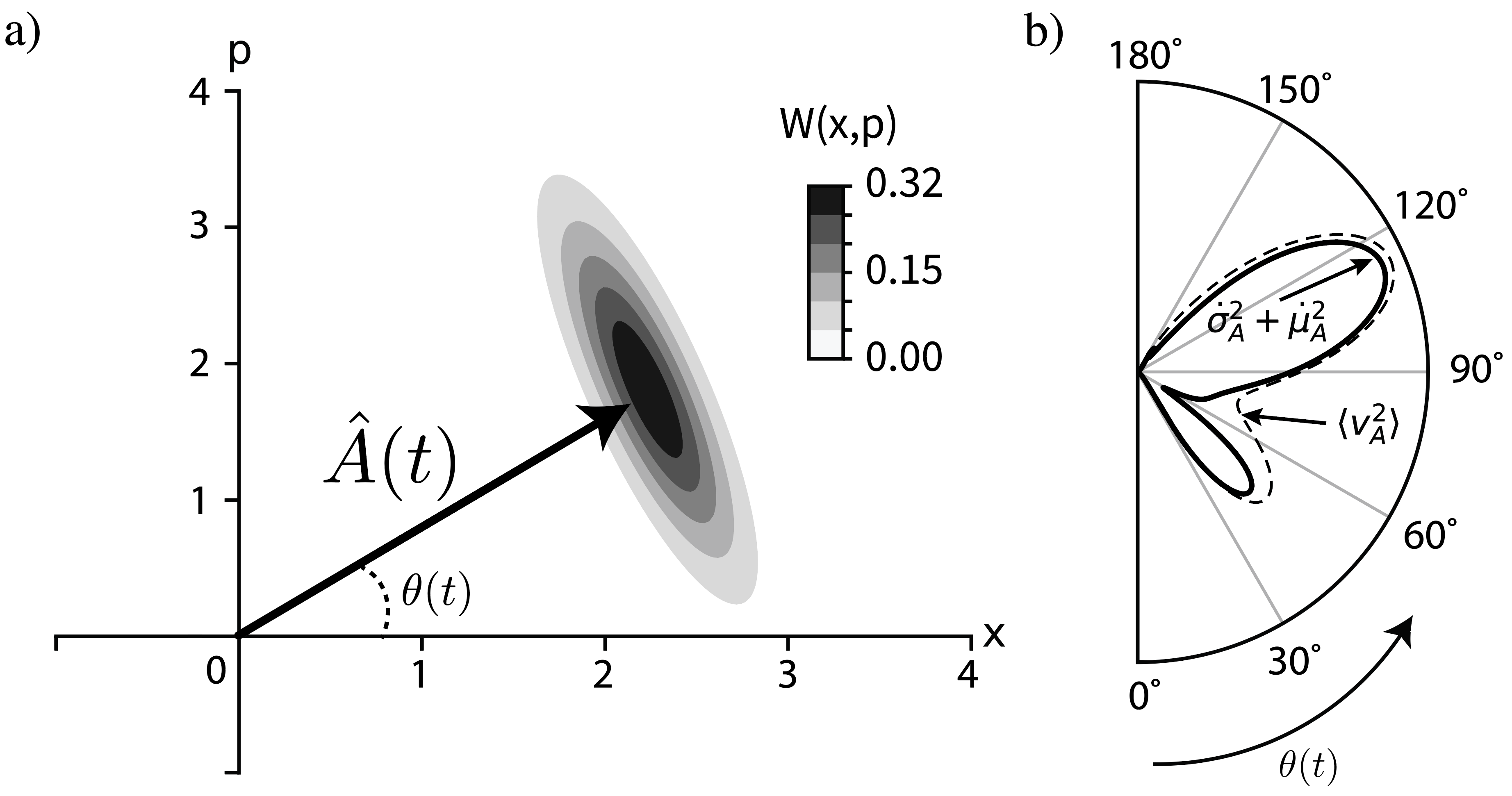}\caption{Numerical verification
of the inequality $(\dot{\mu}_{A})^{2}+\left(  \dot{\sigma}_{A}\right)
^{2}\leq\left\langle v_{A}^{2}\right\rangle $ using a squeezed coherent state
and a time-dependent operator $\hat{A}\left(  t\right)  \overset{\text{def}%
}{=}\cos\left[  \theta\left(  t\right)  \right]  \hat{x}+\sin\left[
\theta\left(  t\right)  \right]  \hat{p}$, where $\theta\left(  t\right)
\overset{\text{def}}{=}\cos\left(  t\right)  $. In a), we plot the Wigner
function $W\left(  x\text{, }p\right)  $ of the squeezed state used for the
initial conditions in the position-momentum space. In addition, we note in a)
that the operator $\hat{A}(t)$ forms an angle $\theta\left(  t\right)  $ with
the position axis. In the example, the squeezed coherent state is
characterized by $\alpha\overset{\text{def}}{=}2+i$, and $z\overset
{\text{def}}{=}0.5+0.5i$. In b), we visualize the inequality $(\dot{\mu}%
_{A})^{2}+\left(  \dot{\sigma}_{A}\right)  ^{2}\leq\left\langle v_{A}%
^{2}\right\rangle $ as a function of $\theta$ in a polar plot. The black solid
line represents $(\dot{\mu}_{A})^{2}+\left(  \dot{\sigma}_{A}\right)  ^{2}$
and is always bounded by the dashed line that describes $\left\langle
v_{A}^{2}\right\rangle $. Thus, the inequality is constantly preserved.
Finally, we assume $\omega=\hslash=m=1$ in all our numerical calculations.}%
\end{figure}\ 

Finally, the validity of the inequality $(\dot{\mu}_{A})^{2}+\left(
\dot{\sigma}_{A}\right)  ^{2}\leq\left\langle v_{A}^{2}\right\rangle $ for a
physical system represented by a quantum harmonic oscillator in a
finite-dimensional Fock space is illustrated in Fig. $3$. The inequality
$(\dot{\mu}_{A})^{2}+\left(  \dot{\sigma}_{A}\right)  ^{2}\leq\left\langle
v_{A}^{2}\right\rangle $ establishes fundamental limits governing the
evolution of the time-dependent homodyne observable $\hat{A}\left(  t\right)
=\cos\left[  \theta\left(  t\right)  \right]  \hat{x}+\sin\left[
\theta\left(  t\right)  \right]  \hat{p}$ for displaced squeezed states. This
mathematical relationship encapsulates profound physical constraints on how
quantum systems can evolve when probed through continuous measurements. The
displacement parameter $\alpha$ introduces classical amplitude to the quantum
state, creating measurable expectation values in the quadrature fields.
Simultaneously, the squeezing parameter $z$ generates anisotropic quantum
noise, redistributing fluctuations between conjugate variables. The inequality
reveals that these two features -- classical displacement and quantum
squeezing -- cannot evolve arbitrarily fast when considered together. On the
left side of the inequality, $\dot{\mu}_{A}$ represents the rate of change of
the mean signal, corresponding to how rapidly the state's centroid moves
through phase space. The $\dot{\sigma}_{A}$ term captures the evolution rate
of quantum fluctuations, describing how quickly the noise profile can
reconfigure. The right side $\left\langle v_{A}^{2}\right\rangle $\textbf{
}represents the maximum possible value for the sum of these squared rates, set
by the system's Hamiltonian. This constraint becomes particularly significant
in experimental quantum optics. When performing homodyne measurements with a
rotating reference phase\textbf{ }$\theta\left(  t\right)  =\cos(t)$, the
inequality determines: (i) how quickly measurement outcomes can track the
system dynamics, and (ii) how rapidly the measurement noise can be optimized.
The bound is approached when the displacement and squeezing become optimally
aligned with the instantaneous measurement quadrature. In practical
applications, this relationship has crucial implications. For quantum control
protocols \cite{dalessandro21}, it sets speed limits for feedback operations
on squeezed states. In quantum metrology \cite{giova11}, it establishes
fundamental trade-offs between measurement bandwidth and precision. The
inequality in Eq. (\ref{eq:29}) essentially quantifies how quantum mechanics
restricts our ability to simultaneously track and control both the mean values
and fluctuations of observables in unitarily evolving quantum systems.

It is worthwhile pointing out that we performed a numerical assessment of our
inequality through the use of a lossless harmonic oscillator. However, it is
important to note that inherent losses are unavoidable in real-world optical
systems, particularly those that incorporate lossy beam splitters
\cite{loudon98}. In general, losses may arise from dispersive ohmic effects or
from the difficulties encountered in managing and capturing light within
dielectric scattering materials. Therefore, it would be valuable to expand our
analysis to incorporate certain loss mechanisms \cite{seni60}. We intend to
investigate this aspect in our upcoming research efforts, which will focus on
nonunitary evolutions of open quantum systems.

\section{Conclusion}

In this paper, we presented an alternative derivation of the fact that, in
unitary quantum dynamics, the speed of the standard deviation of any
observable $A$ is constrained by the standard deviation of its corresponding
velocity-like observable $v_{A}$ (Appendix B). This inequality in Eq.
(\ref{ineq22}), originally derived in Ref. \cite{H2} by Hamazaki, was
recovered here by using previously developed methods for achieving upper
limits on the acceleration in projective Hilbert space of arbitrary
finite-dimensional quantum systems whose dynamics is governed by any
time-dependent Hamiltonian \cite{AC,CB}. In particular, we extended our
results on the acceleration of a quantum evolution in projective Hilbert space
being upper bounded by the standard deviation of the time derivative of the
Hamiltonian, to include any observable $A$ within the framework of unitary
quantum evolution. In the end, we discussed three examples. In the first two
examples, we considered the unitary dynamics of two-level quantum systems
indicating a loose and a tight bound on fluctuation growth of suitably chosen
observables. In our third example, we verified the validity of the
inequalities in Eqs. (\ref{ineq22}) and (\ref{a4}) for a multi--level quantum
system represented by a harmonic oscillator in a finite-dimensional Fock space.

\subsection{Main results}

Our main findings can be outlined as follows:

\begin{enumerate}
\item[{[i]}] We revisited in a quantitative manner Hamazaki's derivation
\cite{H2} of speed limits to fluctuation dynamics restricted to unitary
quantum-mechanical evolutions.

\item[{[ii]}] Following derivations of quantum acceleration limits in
projective Hilbert space starting from conventional quantum-mechanical
uncertainty relations \cite{P,AC,CB}, we presented an alternative derivation
of the fact that the speed of an observable's fluctuation is upper bounded by
the fluctuation of a suitably defined velocity observable (i.e., $\left\vert
d\sigma_{A}/dt\right\vert \leq\sigma_{v_{A}}$). We also pointed that the
inequality can be regarded as expressing the fact that there exists a
trade-off between the speeds of the mean and the standard deviation for
observables in unitary dynamics (i.e., $\left(  d\left\langle A\right\rangle
/dt\right)  ^{2}+\left(  d\sigma_{A}/dt\right)  ^{2}\leq\left\langle v_{A}%
^{2}\right\rangle $).

\item[{[iii]}] We presented illustrative examples limited to the unitary
dynamics of both two-level (Fig\textbf{. }$1$\textbf{) }and multi-level
(Fig\textbf{. }$3$)\textbf{ }quantum\textbf{ }systems where suitably chosen
observables are specified by tight (i.e., $\left(  d\left\langle
A\right\rangle /dt\right)  ^{2}+\left(  d\sigma_{A}/dt\right)  ^{2}%
=\left\langle v_{A}^{2}\right\rangle $) or loose (i.e., $\left(  d\left\langle
A\right\rangle /dt\right)  ^{2}+\left(  d\sigma_{A}/dt\right)  ^{2}%
\leq\left\langle v_{A}^{2}\right\rangle $, Fig. $1$) upper bounds on their
fluctuation growth. Our preliminary analysis indicates that increased
fluctuation rates correlate with diminished relative qualities of the signals
(Fig. $2$ and Appendix D).

\item[{[iv]}] We showed that the inequality in Eq. (\ref{eq:29}) fundamentally
measures the limitations imposed by quantum mechanics on our capacity to
simultaneously monitor and control both the average values and fluctuations of
observables in quantum-mechanical systems that change over time. This was
demonstrated in spin precessions in alternating current magnetic fields
(Examples\textbf{ }$1$\textbf{ }and\textbf{ }$2$\textbf{) }and, in addition,
in quantum optical systems specified by a harmonic oscillator within a
finite-dimensional Fock space\textbf{ (}Example\textbf{ }$3$\textbf{).}
\end{enumerate}

\medskip

\subsection{Discussion and outlook}

From a theoretical perspective, this study holds inherent significance as it
addresses previously unexamined statistical inequalities related to the
dynamic behavior of quantum observables under nonequilibrium conditions. From
a practical standpoint, this investigation can act as a fundamental basis for
developing physically-based figures of merit that quantify experimentally
observable intensity levels of fluctuations within intricate quantum systems
\cite{H1,H2,H3}. Such quantifiers can subsequently facilitate the creation of
quantum control strategies aimed at enhancing the system's dynamics regarding
speed, efficiency, and complexity \cite{carlo25A,carlo25B,carlo25C,carlo25D}.
Nevertheless, at this moment, these observations are largely speculative. We
aim to conduct more detailed quantitative research on these matters in our
future scientific endeavors.

\medskip

We would like to stress that in contrast to Hamazaki's proof, our derivation
is confined to unitary dynamics, as it employs proof techniques that
demonstrate the limits of quantum acceleration. These limits have thus far
been addressed exclusively within the framework of closed quantum systems.
Nevertheless, our derivation provides a lucid explanation that, at a
fundamental level, these upper limits on the growth of observable fluctuations
are fundamentally rooted in the standard uncertainty relations of quantum mechanics.

\medskip

However, our different proof together with clear illustrative examples
pertinent to quantum information science, results in an inequality that
fundamentally evaluates the extent to which quantum mechanics restricts our
ability to simultaneously observe and control both the mean values and
fluctuations of observables in quantum systems that evolve unitarily. This
restriction may, in turn, create opportunities for new and significant lines
of inquiry within the ever-growing field of quantum fluctuations and
uncertainty relations in nonequilibrium thermodynamics. Despite the formal
mathematical similarities, our proof is distinctly different from a physics
perspective when compared to the one in Ref. \cite{H2} (see Table I).
Specifically, while Hamazaki's proof is broad in its scope and is based on
general statistical equalities that are applicable in both classical and
quantum physical frameworks with a robust underlying probabilistic structure,
our proof is more narrowly focused and specifically relies on the algebra of
observables in quantum mechanics. Interestingly, the proof derived from the
algebra of operators provides opportunities to identify in a schematic way the
physical configurations for which one can achieve tight upper bounds instead
of loose upper bounds. Lastly, since the extension of Hamazaki's analysis to
open quantum systems exhibits unresolved issues \cite{H2}, our approach can
pave the way to alternative perspectives that can help generalizing these
inequalities to more realistic quantum systems.

\medskip

Real physical systems are predominantly open and engage in interactions with
an external environment or bath \cite{taddei13,deffner13,okuyama18,campo13}.
Such interactions typically lead to dissipation within the system. It would be
valuable to investigate the temporal dynamics of expectation values and
variances of observables in dissipative quantum systems \cite{darius12}, such
as damped harmonic oscillators. Transitioning from unitary (closed) to
nonunitary (open) quantum mechanical evolutions presents several unresolved
challenges, including the need for a suitable definition of a velocity
observable \cite{H2} and the appropriate management of uncertainty relations
for quantum systems existing in mixed states \cite{luo05,ole14}.

\medskip

As our final remark, we point out that it is known that there exists a strong
relationship between mean and variance changes (i.e., $d\mu_{A}/dt$ and
$d\sigma_{A}/dt$, respectively) in several fields of science, including
climate change scenarios \cite{mearns97}. It is also acknowledged that lower
\cite{barato15,horowitz20} and upper \cite{bakewell23} bounds on the size of
fluctuations of dynamical observables are very important since having both of
them is necessary to limit the range of estimation errors. In these bounds,
the main quantity of interest is the so-called ratio of variance to mean (or,
alternatively, the squared relative uncertainty of the observable $A$)
$\varepsilon_{A}$ with $\varepsilon_{A}^{2}\overset{\text{def}}{=}%
\mathrm{var}(A)/\left\langle A\right\rangle ^{2}=\sigma_{A}^{2}/\mu_{A}^{2}$.
In particular, an uncertainty $\varepsilon_{A}$ requires at least a
(thermodynamic) cost of $2k_{B}T/\varepsilon_{A}^{2}=T\sigma t$. Here, $k_{B}$
is the Boltzmann constant, $\sigma t$ is the average entropy produced in a
time interval $t$, and $\sigma$ denotes a constant entropy production rate for
a stochastic (classical) dynamical systems in an out-of-equilibrium
configuration in which it dissipates energy towards an external environment at
fixed temperature $T$. We note that the time-derivative of the $\varepsilon
_{A}^{2}$ not only depends on $\mu_{A}$ and $\sigma_{A}$, it is also a
function of the rates of change $d\mu_{A}/dt$ and $d\sigma_{A}/dt$ since
$d\varepsilon_{A}^{2}/dt=2\left(  \sigma_{A}/\mu_{A}^{3}\right)  \left[
\mu_{A}\left(  d\sigma_{A}/dt\right)  -\sigma_{A}\left(  d\mu_{A}/dt\right)
\right]  $. From this latter equation, we clearly see that both $d\mu_{A}/dt$
and $d\sigma_{A}/dt$ play an essential role in specifying $d\varepsilon
_{A}^{2}/dt$. For this reason, it would be interesting to understand how an
upper bound on $\left(  d\mu_{A}/dt\right)  ^{2}+\left(  d\sigma
_{A}/dt\right)  ^{2}$ would help in constraining the rate of change of the
squared uncertainty $\varepsilon_{A}^{2}$ in out-of-equilibrium dynamical
situations, possibly fully quantum \cite{carollo19,girotti23}. From an
experimental standpoint, it would be interesting to verify the validity of
inequalities in Eqs. (\ref{ineq22}) and (\ref{a2}). For the reader interested
in how to experimentally measure the mean and the variance of
quantum-mechanical observables, we suggest
Ref.\cite{andersson07,stano08,wu09,piccirillo15,luo20,ogawa21}. We leave these
intriguing points to future investigations.

\medskip

In summary, notwithstanding the existing constraints, we are strongly
persuaded that our research will inspire additional scholars and facilitate
further in-depth explorations into the connections between uncertainty
relations, quantum acceleration limits, and ultimately, the growth of
fluctuations in observables within intricate quantum dynamical contexts.

\begin{acknowledgments}
Any opinions, findings and conclusions or recommendations expressed in this
material are those of the author(s) and do not necessarily reflect the views
of their home Institutions. The authors thank two anonymous referees for very
useful comments leading to an improved version of this manuscript.
\end{acknowledgments}

\bigskip\pagebreak

\appendix

\section{Deriving the Mandelstam-Tamm Bound}

In this appendix, we revisit the Mandelstam-Tamm derivation of the minimum
time for the evolution to an orthogonal state as presented in Ref. \cite{MT}.

Consider a quantum state whose dynamics is governed by the Schr\"{o}dinger
equation, where any observable $A$ that is not explicitly time-dependent
(i.e., such that $\partial A/\partial t=0$) satisfies the
Liouville-von-Neumann relation%
\begin{equation}
\frac{dA}{dt}=\frac{i}{\hslash}\left[  \mathrm{H}\text{, }A\right]  \text{.}
\label{aa1}%
\end{equation}
Following the notation used in Ref. \cite{MT}, we recall that the generalized
Robertson uncertainty relation for any two operators $A$ and $B$ implies that
$\left(  \Delta A\right)  \left(  \Delta B\right)  \geq\left\vert \left\langle
\left[  A\text{, }B\right]  \right\rangle \right\vert /2$, with $\Delta
A\overset{\text{def}}{=}\sqrt{\left\langle A^{2}\right\rangle -\left\langle
A\right\rangle ^{2}}$. As a side remark, we stress that while $\Delta A$ is a
positive scalar quantity that specifies the standard deviation of the operator
$A$ in this Appendix, $\Delta A\overset{\text{def}}{=}A-\left\langle
A\right\rangle $ denotes an operator in the alternative proof presented in
Section III. Returning to our revisitation, we note that when $B=\mathrm{H}$,
Eq. (\ref{aa1}) together with the Roberston relation yield%
\begin{equation}
\left(  \Delta\mathrm{H}\right)  \left(  \Delta A\right)  \geq\frac{\hslash
}{2}\left\vert \langle\frac{dA}{dt}\rangle\right\vert \text{.} \label{aa2}%
\end{equation}
Assuming that $A\overset{\text{def}}{=}\left\vert \psi\left(  0\right)
\right\rangle \left\langle \psi\left(  0\right)  \right\vert $ is the
projector onto the initial state $\left\vert \psi\left(  0\right)
\right\rangle $, we have $A^{2}=A$ and, thus, $\Delta A\overset{\text{def}}%
{=}\sqrt{\left\langle A^{2}\right\rangle -\left\langle A\right\rangle ^{2}%
}=\sqrt{\left\langle A\right\rangle -\left\langle A\right\rangle ^{2}}$.
Therefore, Eq. (\ref{aa2}) reduces to%
\begin{equation}
\frac{\Delta\mathrm{H}}{\hslash}dt\geq-\frac{1}{2}\frac{d\left\langle
A\right\rangle }{\sqrt{\left\langle A\right\rangle -\left\langle
A\right\rangle ^{2}}}\text{,} \label{aa3}%
\end{equation}
since $\left\langle dA/dt\right\rangle =d\left\langle A\right\rangle /dt$ and
$\left\langle A\right\rangle $ decreases in time (with its maximum being $1$
at $t=0$ since $\left\langle A\right\rangle _{0}=1$). Integration of Eq.
(\ref{aa3}) from $0$ to $\Delta T$ leads to%
\begin{equation}
\int_{0}^{\Delta T}\frac{\Delta\mathrm{H}}{\hslash}dt\geq-\frac{1}{2}%
\int_{\left\langle A\right\rangle _{0}}^{\left\langle A\right\rangle _{\Delta
T}}\frac{d\left\langle A\right\rangle }{\sqrt{\left\langle A\right\rangle
-\left\langle A\right\rangle ^{2}}}\text{,} \label{aa4}%
\end{equation}
that is%
\begin{equation}
\frac{1}{\hslash}\left(  \Delta\mathrm{H}\right)  \left(  \Delta T\right)
\geq\frac{\pi}{2}-\arcsin\left[  \left\vert \left\langle \psi\left(  0\right)
\left\vert \psi\left(  \Delta T\right)  \right.  \right\rangle \right\vert
\right]  \text{,}%
\end{equation}
given that $\Delta\mathrm{H}$ in Eq. (\ref{aa4}) is assumed to be
time-independent. In particular, if the initial and final states are assumed
to be orthogonal, we finally arrive at the MT bound%
\begin{equation}
\left(  \Delta\mathrm{H}\right)  \left(  \Delta T\right)  \geq\frac{h}%
{4}\text{.} \label{aa5}%
\end{equation}
The derivation of the inequality in (\ref{aa5}) ends our presentation here.
For further details on alternative derivations of the MT\ bound, we refer to
Refs. \cite{anandan90,vaidman92,deffner17,cafaro21,ole22,giorgi24}.

\section{Defining the Velocity Observable}

In this appendix, we explain the meaning of the concept of velocity observable
in unitary quantum dynamics as originally presented by Hamazaki in Ref.
\cite{H2}.

Hamazaki defines the velocity observable $v_{A}$ as%
\begin{equation}
v_{A}\overset{\text{def}}{=}\frac{i}{\hslash}\left[  \mathrm{H}\text{,
}A\right]  +\dot{A}\text{,} \label{1}%
\end{equation}
with $\dot{A}\overset{\text{def}}{=}\partial A/\partial t$. Moreover, he
interprets $v_{A}$ in Eq. (\ref{1}) as the Schr\"{o}dinger picture of the time
derivative of the observable in the Heisenberg representation. To understand
this interpretation, we begin by discussing the Heisenberg and the
Schr\"{o}dinger of quantum observables, respectively.

\subsection{The Heisenberg representation}

In quantum mechanics \cite{sakurai20}, the expectation value of an observable
$A$ can be expressed in two alternative manners as%
\begin{equation}
\left\langle A\right\rangle =\left\langle \psi\left(  0\right)  \left\vert
A_{H}\left(  t\right)  \right\vert \psi\left(  0\right)  \right\rangle \text{,
or }\left\langle A\right\rangle =\left\langle \psi\left(  t\right)  \left\vert
A_{S}\left(  t\right)  \right\vert \psi\left(  t\right)  \right\rangle
\text{,}%
\end{equation}
with $A_{H}\left(  t\right)  \overset{\text{def}}{=}U^{\dagger}\left(
t\right)  A_{S}\left(  t\right)  U\left(  t\right)  $ being the observable $A$
in the Heisenberg representation, $A_{S}\left(  t\right)  $ is the observable
$A$ in the Schr\"{o}dinger representation, and $U\left(  t\right)  $ is the
unitary evolution operator satisfying the relation $i\hslash\partial
_{t}U\left(  t\right)  =\mathrm{H}_{S}\left(  t\right)  U\left(  t\right)  $
with $\mathrm{H}_{S}\left(  t\right)  $ denoting the Hamiltonian of the system
in the Schr\"{o}dinger representation. From $\left\langle A\right\rangle
=\left\langle \psi\left(  0\right)  \left\vert A_{H}\left(  t\right)
\right\vert \psi\left(  0\right)  \right\rangle $, we get%
\begin{equation}
\frac{d\left\langle A\right\rangle }{dt}=\frac{d\left\langle \psi\left(
0\right)  \left\vert A_{H}\left(  t\right)  \right\vert \psi\left(  0\right)
\right\rangle }{dt}=\langle\psi\left(  0\right)  \left\vert \frac
{dA_{H}\left(  t\right)  }{dt}\right\vert \psi\left(  0\right)  \rangle
=\langle\frac{dA_{H}}{dt}\rangle\text{,}%
\end{equation}
that is,%
\begin{equation}
\frac{d\left\langle A\right\rangle }{dt}=\langle\frac{dA_{H}}{dt}%
\rangle\text{.} \label{2}%
\end{equation}
For completeness, let us find an explicit expression for $dA_{H}/dt$ in Eq.
(\ref{2}). We have,%
\begin{align}
\frac{dA_{H}\left(  t\right)  }{dt}  &  =\frac{d}{dt}\left[  U^{\dagger
}\left(  t\right)  A_{S}\left(  t\right)  U\left(  t\right)  \right]
\nonumber\\
&  =\frac{\partial U^{\dagger}\left(  t\right)  }{\partial t}A_{S}\left(
t\right)  U\left(  t\right)  +U^{\dagger}\left(  t\right)  \frac{\partial
A_{S}\left(  t\right)  }{\partial t}U\left(  t\right)  +U^{\dagger}\left(
t\right)  A_{S}\left(  t\right)  \frac{\partial U(t)}{\partial t}\nonumber\\
&  =-\frac{1}{i\hslash}U^{\dagger}\left(  t\right)  \mathrm{H}_{S}\left(
t\right)  A_{S}\left(  t\right)  U\left(  t\right)  +U^{\dagger}\left(
t\right)  \frac{\partial A_{S}\left(  t\right)  }{\partial t}U\left(
t\right)  +\frac{1}{i\hslash}U^{\dagger}\left(  t\right)  A_{S}\left(
t\right)  \mathrm{H}_{S}\left(  t\right)  U\left(  t\right) \nonumber\\
&  =-\frac{1}{i\hslash}U^{\dagger}\left(  t\right)  \mathrm{H}_{S}\left(
t\right)  U\left(  t\right)  U^{\dagger}\left(  t\right)  A_{S}\left(
t\right)  U\left(  t\right)  +U^{\dagger}\left(  t\right)  \frac{\partial
A_{S}\left(  t\right)  }{\partial t}U\left(  t\right)  +\frac{1}{i\hslash
}U^{\dagger}\left(  t\right)  A_{S}\left(  t\right)  U\left(  t\right)
U^{\dagger}\left(  t\right)  \mathrm{H}_{S}\left(  t\right)  U\left(  t\right)
\nonumber\\
&  =-\frac{1}{i\hslash}\mathrm{H}_{H}\left(  t\right)  A_{H}\left(  t\right)
+U^{\dagger}\left(  t\right)  \frac{\partial A_{S}\left(  t\right)  }{\partial
t}U\left(  t\right)  +\frac{1}{i\hslash}A_{H}\left(  t\right)  \mathrm{H}%
_{H}\left(  t\right) \nonumber\\
&  =\left(  \frac{\partial A_{S}\left(  t\right)  }{\partial t}\right)
_{H}+\frac{1}{i\hslash}\left[  A_{H}\left(  t\right)  \text{, }\mathrm{H}%
_{H}\left(  t\right)  \right]  \text{,}%
\end{align}
that is,%
\begin{equation}
\frac{dA_{H}\left(  t\right)  }{dt}=\left(  \frac{\partial A_{S}\left(
t\right)  }{\partial t}\right)  _{H}+\frac{1}{i\hslash}\left[  A_{H}\left(
t\right)  \text{, }\mathrm{H}_{H}\left(  t\right)  \right]  \text{,}
\label{berry}%
\end{equation}
where $\mathrm{H}_{H}\left(  t\right)  \overset{\text{def}}{=}U^{\dagger
}\left(  t\right)  \mathrm{H}_{S}\left(  t\right)  U\left(  t\right)  $ is the
Hamiltonian of the system in the Heisenberg representation, and
\begin{equation}
\left(  \frac{\partial A_{S}\left(  t\right)  }{\partial t}\right)
_{H}\overset{\text{def}}{=}U^{\dagger}\left(  t\right)  \frac{\partial
A_{S}\left(  t\right)  }{\partial t}U\left(  t\right)  \text{.}%
\end{equation}
As a side remark, we note that if $\mathrm{H}_{S}$ is time-independent and
equals \textrm{H}, then $\mathrm{H}_{H}\equiv\mathrm{H}_{S}\equiv\mathrm{H}$.
Then,\ Eq. (\ref{berry}) reduces to%
\begin{equation}
\frac{dA_{H}\left(  t\right)  }{dt}=e^{\frac{i}{\hslash}\mathrm{H}t}%
\frac{\partial A_{S}\left(  t\right)  }{\partial t}e^{-\frac{i}{\hslash
}\mathrm{H}t}+\frac{1}{i\hslash}\left[  A_{H}\left(  t\right)  \text{,
}\mathrm{H}\right]  \text{.} \label{berry2}%
\end{equation}
Moreover, if $A_{S}$ is time-independent, then%
\begin{equation}
\frac{dA_{H}\left(  t\right)  }{dt}=\frac{1}{i\hslash}\left[  A_{H}\left(
t\right)  \text{, }\mathrm{H}\right]  \text{.}%
\end{equation}
In summary, in the Heisenberg representation, one can set in the most general
case that%
\begin{equation}
v_{A}^{H}\overset{\text{def}}{=}\frac{dA_{H}\left(  t\right)  }{dt}=\left(
\frac{\partial A_{S}\left(  t\right)  }{\partial t}\right)  _{H}+\frac
{1}{i\hslash}\left[  A_{H}\left(  t\right)  \text{, }\mathrm{H}_{H}\left(
t\right)  \right]  \text{, with }\left\langle v_{A}^{H}\right\rangle
=\frac{d\left\langle A\right\rangle }{dt}=\langle\frac{dA_{H}}{dt}%
\rangle\text{.}%
\end{equation}
We are now ready to discuss the concept of time-derivative of a quantum
observable in the Schr\"{o}dinger representation.

\subsection{The Schr\"{o}dinger representation}

In the Schr\"{o}dinger representation, we have $\left\langle A\right\rangle
=\left\langle \psi\left(  t\right)  \left\vert A_{S}\left(  t\right)
\right\vert \psi\left(  t\right)  \right\rangle $. Therefore, we get%
\begin{align}
\frac{d\left\langle A\right\rangle }{dt}  &  =\frac{d\left\langle \psi\left(
t\right)  \left\vert A_{S}\left(  t\right)  \right\vert \psi\left(  t\right)
\right\rangle }{dt}\nonumber\\
&  =\left\langle \dot{\psi}\left(  t\right)  \left\vert A_{S}\left(  t\right)
\right\vert \psi\left(  t\right)  \right\rangle +\langle\psi\left(  t\right)
\left\vert \frac{\partial A_{S}\left(  t\right)  }{\partial t}\right\vert
\psi\left(  t\right)  \rangle+\left\langle \psi\left(  t\right)  \left\vert
A_{S}\left(  t\right)  \right\vert \dot{\psi}\left(  t\right)  \right\rangle
\nonumber\\
&  =-\frac{1}{i\hslash}\left\langle \psi\left(  t\right)  \left\vert
\mathrm{H}_{S}\left(  t\right)  A_{S}\left(  t\right)  \right\vert \psi\left(
t\right)  \right\rangle +\langle\psi\left(  t\right)  \left\vert
\frac{\partial A_{S}\left(  t\right)  }{\partial t}\right\vert \psi\left(
t\right)  \rangle+\frac{1}{i\hslash}\left\langle \psi\left(  t\right)
\left\vert A_{S}\left(  t\right)  \mathrm{H}_{S}\left(  t\right)  \right\vert
\psi\left(  t\right)  \right\rangle \nonumber\\
&  =\langle\psi\left(  t\right)  \left\vert \frac{\partial A_{S}\left(
t\right)  }{\partial t}\right\vert \psi\left(  t\right)  \rangle+\frac
{1}{i\hslash}\left\langle \psi\left(  t\right)  \left\vert \left[
A_{S}\left(  t\right)  \text{, }\mathrm{H}_{S}\left(  t\right)  \right]
\right\vert \psi\left(  t\right)  \right\rangle \nonumber\\
&  =\langle\psi\left(  t\right)  \left\vert \frac{\partial A_{S}\left(
t\right)  }{\partial t}+\frac{1}{i\hslash}\left[  A_{S}\left(  t\right)
\text{, }\mathrm{H}_{S}\left(  t\right)  \right]  \right\vert \psi\left(
t\right)  \rangle\nonumber\\
&  =\langle\psi\left(  t\right)  \left\vert \frac{dA_{S}\left(  t\right)
}{dt}\right\vert \psi\left(  t\right)  \rangle\nonumber\\
&  =\left\langle \psi\left(  t\right)  \left\vert v_{A}^{S}\right\vert
\psi\left(  t\right)  \right\rangle \nonumber\\
&  =\left\langle v_{A}^{S}\right\rangle \text{.}%
\end{align}
In summary, in the Schr\"{o}dinger representation, one can set in the most
general case that%
\begin{equation}
v_{A}^{S}\overset{\text{def}}{=}\frac{dA_{S}\left(  t\right)  }{dt}%
=\frac{\partial A_{S}\left(  t\right)  }{\partial t}+\frac{1}{i\hslash}\left[
A_{S}\left(  t\right)  \text{, }\mathrm{H}_{S}\left(  t\right)  \right]
\text{, with }\left\langle v_{A}^{S}\right\rangle =\frac{d\left\langle
A\right\rangle }{dt}=\langle\frac{dA_{S}}{dt}\rangle\text{.}%
\end{equation}
We stress that $dA_{S}/dt$ is used here as a notation for $v_{A}^{S}$ and
equals $\partial A_{S}\left(  t\right)  /\partial t+\left(  i\hslash\right)
^{-1}\left[  A_{S}\left(  t\right)  \text{, }\mathrm{H}_{S}\left(  t\right)
\right]  $. What is the relation between $v_{A}^{S}$ and $v_{A}^{H}$? We
observe that,%
\begin{align}
v_{A}^{H}  &  =\frac{dA_{H}\left(  t\right)  }{dt}\nonumber\\
&  =\left(  \frac{\partial A_{S}\left(  t\right)  }{\partial t}\right)
_{H}+\frac{1}{i\hslash}\left[  A_{H}\left(  t\right)  \text{, }\mathrm{H}%
_{H}\left(  t\right)  \right] \nonumber\\
&  =U^{\dagger}\left(  t\right)  \frac{\partial A_{S}\left(  t\right)
}{\partial t}U\left(  t\right)  +\frac{1}{i\hslash}\left[  U^{\dagger}\left(
t\right)  \mathrm{A}_{S}\left(  t\right)  U\left(  t\right)  \text{,
}U^{\dagger}\left(  t\right)  \mathrm{H}_{S}\left(  t\right)  U\left(
t\right)  \right] \nonumber\\
&  =U^{\dagger}\left(  t\right)  \left(  \frac{\partial A_{S}\left(  t\right)
}{\partial t}+\frac{1}{i\hslash}\left[  A_{S}\left(  t\right)  \text{,
}\mathrm{H}_{S}\left(  t\right)  \right]  \right)  U\left(  t\right)
\nonumber\\
&  =U^{\dagger}\left(  t\right)  v_{A}^{S}U\left(  t\right)  \text{,}%
\end{align}
that is,%
\begin{equation}
v_{A}^{H}=U^{\dagger}\left(  t\right)  v_{A}^{S}U\left(  t\right)  \text{,}
\label{impo}%
\end{equation}
with $v_{A}^{S}\overset{\text{def}}{=}\partial_{t}A_{S}\left(  t\right)
+\left(  i\hslash\right)  ^{-1}\left[  A_{S}\left(  t\right)  \text{,
}\mathrm{H}_{S}\left(  t\right)  \right]  $. From Eq. (\ref{impo}), we can
understand that $v_{A}^{S}$ denotes the Schr\"{o}dinger picture of $v_{A}^{H}$
(i.e., the time derivative of the observable in the Heisenberg
representation). In summary, considering the correspondence between Hamazaki's
notation and ours, we have%
\begin{equation}
\left(  v_{A}\right)  _{\mathrm{Hamazaki}}\rightarrow v_{A}^{S}\text{,
}\left(  \frac{i}{\hslash}\left[  \mathrm{H}\text{, }A\right]  \right)
_{\mathrm{Hamazaki}}\rightarrow\frac{1}{i\hslash}\left[  A_{S}\left(
t\right)  \text{, }\mathrm{H}_{S}\left(  t\right)  \right]  \text{, and
}\left(  \overset{.}{A}\right)  _{\mathrm{Hamazaki}}\rightarrow\frac{\partial
A_{S}\left(  t\right)  }{\partial t}\text{.}%
\end{equation}
In general, one does not use the cumbersome notation $A_{H}\left(  t\right)  $
and $A_{S}\left(  t\right)  $. One simply writes $A(t)=U^{\dagger}\left(
t\right)  AU\left(  t\right)  $ and $A$, respectively. Furthermore, although
$v_{A}^{H}$ is generally different from $v_{A}^{S}$, we have $\langle
v_{A}^{H}\rangle_{\left\vert \psi\left(  0\right)  \right\rangle }=\langle
v_{A}^{S}\rangle_{\left\vert \psi(t)\right\rangle }=\left\langle
v_{A}\right\rangle $. We also recognize that we can replace $\left\langle
\cdot\right\rangle _{\left\vert \psi\left(  0\right)  \right\rangle }$ and
$\left\langle \cdot\right\rangle _{\left\vert \psi\left(  t\right)
\right\rangle }$ with simply $\left\langle \cdot\right\rangle $, if we keep in
mind that expectation values of observables in the Heisenberg and
Schr\"{o}dinger representations are evaluated with respect to $\left\vert
\psi\left(  0\right)  \right\rangle $ and $\left\vert \psi\left(  t\right)
\right\rangle $, respectively. Indeed, this notation was adopted throughout
this work. With this remark, we end our discussion here on the concept of
velocity observable in unitary quantum dynamics.

\section{Geometry of tight upper bounds}

In this appendix, following our remarks in Section IV, we discuss possible
geometric configurations for two-level quantum systems that yield to tight
bounds when considering the inequality,%
\begin{equation}
\frac{\left[  \mathbf{m\cdot\dot{m}-}\left(  \mathbf{a\cdot m}\right)  \left(
\mathbf{a\cdot\dot{m}}\right)  \right]  ^{2}}{\mathbf{m}^{2}-\left(
\mathbf{a\cdot m}\right)  ^{2}}\leq\left[  \mathbf{\dot{m}+}2\left(
\mathbf{m\times h}\right)  \right]  ^{2}-\left\{  \mathbf{a\cdot}\left[
\mathbf{\dot{m}+}2\left(  \mathbf{m\times h}\right)  \right]  \right\}
^{2}\text{.} \label{feel1}%
\end{equation}
We begin by noticing that the numerator of the LHS of Eq. (\ref{feel1}) can be
rewritten as%
\begin{equation}
\left[  \mathbf{m\cdot\dot{m}-}\left(  \mathbf{a\cdot m}\right)  \left(
\mathbf{a\cdot\dot{m}}\right)  \right]  ^{2}=\left\{  \mathbf{\dot{m}\cdot
}\left[  \mathbf{m-}\left(  \mathbf{a\cdot m}\right)  \mathbf{a}\right]
\right\}  ^{2}=\mathbf{\dot{m}\cdot m}_{\perp}\text{,}%
\end{equation}
with $\mathbf{m}_{\perp}\overset{\text{def}}{=}\mathbf{m-}\left(
\mathbf{a\cdot m}\right)  \mathbf{a}$ being the component of $\mathbf{m}$
orthogonal to $\mathbf{a}$. We also have that the denominator of the LHS of
Eq. (\ref{feel1}) becomes $\mathbf{m}_{\perp}\cdot\mathbf{m}_{\perp
}=\mathbf{m}^{2}-\left(  \mathbf{a\cdot m}\right)  ^{2}$. Therefore, the
LHS\ of Eq. (\ref{feel1}) reduces to%
\begin{equation}
\text{LHS}=\frac{\left(  \mathbf{\dot{m}\cdot m}_{\perp}\right)  ^{2}%
}{\mathbf{m}_{\perp}\cdot\mathbf{m}_{\perp}}=\left\Vert \text{\textrm{Proj}%
}_{\mathbf{m}_{\perp}}(\mathbf{\dot{m}})\right\Vert ^{2}\text{,} \label{feel2}%
\end{equation}
with \textrm{Proj}$_{\mathbf{m}_{\perp}}(\mathbf{\dot{m}})$ being the
projection of $\mathbf{\dot{m}}$ onto $\mathbf{m}_{\perp}$ defined as%
\begin{equation}
\text{\textrm{Proj}}_{\mathbf{m}_{\perp}}(\mathbf{\dot{m}})\overset
{\text{def}}{=}\frac{\mathbf{\dot{m}\cdot m}_{\perp}}{\mathbf{m}_{\perp}%
\cdot\mathbf{m}_{\perp}}\mathbf{m}_{\perp}\text{.}%
\end{equation}
Eq. (\ref{feel2}) implies that the LHS of Eq. (\ref{feel1}) is the square of
the projection of $\mathbf{\dot{m}}$ onto the part of $\mathbf{m}$ orthogonal
to $\mathbf{a}$ (i.e., $\mathbf{m}_{\perp}$). Furthermore, focusing on the RHS
of Eq. (\ref{feel1}), we notice that%
\begin{align}
\text{RHS}  &  =\left\Vert \left[  \mathbf{\dot{m}+}2\left(  \mathbf{m\times
h}\right)  \right]  -\left\{  \mathbf{a\cdot}\left[  \mathbf{\dot{m}+}2\left(
\mathbf{m\times h}\right)  \right]  \right\}  \mathbf{a}\right\Vert
^{2}\nonumber\\
&  =\left\Vert \left[  \mathbf{\dot{m}+}2\left(  \mathbf{m\times h}\right)
\right]  _{\bot}\right\Vert ^{2}\text{,} \label{feel3}%
\end{align}
that is, RHS is the square of the magnitude of the vectorial component of
$\mathbf{\dot{m}+}2\left(  \mathbf{m\times h}\right)  $ orthogonal to
$\mathbf{a}$. Then, combining Eqs. (\ref{feel2}) and (\ref{feel3}), the
inequality in Eq. (\ref{feel1}) can be recast as
\begin{equation}
\left\Vert \text{\textrm{Proj}}_{\mathbf{m}_{\perp}}(\mathbf{\dot{m}%
})\right\Vert ^{2}\leq\left\Vert \left[  \mathbf{\dot{m}+}2\left(
\mathbf{m\times h}\right)  \right]  _{\bot}\right\Vert ^{2}\text{.}
\label{feel4}%
\end{equation}
Inspection of Eq. (\ref{feel4}) leads to conclude that, in general, the
equality occurs when
\begin{equation}
\text{\textrm{Proj}}_{\mathbf{a}^{\perp}}(\mathbf{\dot{m}})\propto
\text{\textrm{Proj}}_{\mathbf{a}^{\perp}}\left(  \mathbf{\dot{m}+}2\left(
\mathbf{m\times h}\right)  \right)  \text{,}%
\end{equation}
that is, when there is some coefficient $\lambda\in%
\mathbb{R}
$ for which%
\begin{equation}
\text{\textrm{Proj}}_{\mathbf{a}^{\perp}}(\mathbf{\dot{m}})=\lambda
\text{\textrm{Proj}}_{\mathbf{a}^{\perp}}\left(  \mathbf{m\times h}\right)
\text{,} \label{feel5}%
\end{equation}
with $\mathbf{a}^{\perp}\in\left[  \mathrm{Span}\left\{  \mathbf{a}\right\}
\right]  ^{\perp}$ being an element of the orthogonal complement of the space
spanned by $\mathbf{a}$. Eq. (\ref{feel5}) implies that the equality in Eq.
(\ref{feel4}) is achieved if and only if the projection of $\mathbf{\dot{m}}$
onto the subspace orthogonal to $\mathbf{a}$ is collinear with the projection
of $\mathbf{m\times h}$ onto the same subspace. Then, Eq. (\ref{feel5})
implies that%
\begin{equation}
\left[  \mathbf{\dot{m}-}\lambda\left(  \mathbf{m\times h}\right)  \right]
\cdot\mathbf{v=}0\text{, } \label{feel6}%
\end{equation}
for any $\mathbf{v\perp a}$ with $\mathbf{v\in}\left[  \mathrm{Span}\left\{
\mathbf{a}\right\}  \right]  ^{\perp}$. Finally, Eq. (\ref{feel6}) requires
that $\mathbf{\dot{m}-}\lambda\left(  \mathbf{m\times h}\right)
\in\mathrm{Span}\left\{  \mathbf{a}\right\}  $, that is, $\mathbf{\dot{m}%
=}\lambda\left(  \mathbf{m\times h}\right)  +\mu\mathbf{a}$ for scalar
functions $\lambda$, $\mu\in%
\mathbb{R}
$. Clearly, this leads to the conclusion that a tight bound in Eq.
(\ref{feel1}) can be achieved when $\mathbf{\dot{m}\in}\mathrm{Span}\left\{
\mathbf{m\times h}\text{, }\mathbf{a}\right\}  $. Note that the scalar
functions need not be constant and, in general, they are time-dependent
functions. Interestingly, in our first (second) two-level quantum system
example, we have a tight (loose) bound and $\mathbf{\dot{m}}$ is (is not) an
element of $\mathrm{Span}\left\{  \mathbf{m\times h}\text{, }\mathbf{a}%
\right\}  $.

For a more thorough mathematical analysis accompanied by a perceptive physical
interpretation of these configurations, we direct readers to forthcoming
scientific pursuits. With this last remark, we end our discussion here.

\section{Linking \textrm{SNR}$\left(  A\right)  $ to $\langle v_{A}^{2}%
\rangle$}

In this appendix, we provide a connection between\textbf{ }the \textrm{SNR}%
$\left(  A\right)  \overset{\text{def}}{=}\mu_{A}^{2}/\mathrm{var}(A)$ and
$\left\langle v_{A}^{2}\right\rangle $, where $\left(  d\mu_{A}/dt\right)
^{2}+\left(  d\sigma_{A}/dt\right)  ^{2}\leq\left\langle v_{A}^{2}%
\right\rangle $.

From the inequality $\left(  d\sigma_{A}/dt\right)  ^{2}\leq\sigma_{v_{A}}%
^{2}$, we get that\textbf{ }$-\sigma_{v_{A}}\leq d\sigma_{A}/dt\leq
\sigma_{v_{A}}$. Then, integrating both sides of the inequality $d\sigma
_{A}/dt\leq\sigma_{v_{A}}$ from $0$ to $t$, we obtain%
\begin{equation}
\sigma_{A}\left(  t\right)  \leq\sigma_{A}\left(  0\right)  +\int_{0}%
^{t}\sigma_{v_{A}}\left(  t^{\prime}\right)  dt^{\prime}\text{.} \label{ac1}%
\end{equation}
Noting that both sides in Eq. (\ref{ac1}) are positive and recalling that
$\sigma_{v_{A}}\left(  t^{\prime}\right)  \overset{\text{def}}{=}%
\sqrt{\left\langle v_{A}^{2}\right\rangle \left(  t^{\prime}\right)  -\left[
d\mu_{A}\left(  t^{\prime}\right)  /dt^{\prime}\right]  ^{2}}$, simple
algebraic manipulations lead to%
\begin{equation}
\mathrm{SNR}\left(  A\right)  \geq\left[  \mathrm{SNR}\left(  A\right)
\right]  _{\min} \label{ac2}%
\end{equation}
where $\left[  \mathrm{SNR}\left(  A\right)  \right]  _{\min}$ in Eq.
(\ref{ac2}) is defined as%
\begin{equation}
\left[  \mathrm{SNR}\left(  A\right)  \right]  _{\min}\overset{\text{def}}%
{=}\frac{\mu_{A}^{2}(t)}{\left[  \sigma_{A}\left(  0\right)  +\int_{0}%
^{t}\sqrt{\left\langle v_{A}^{2}\right\rangle \left(  t^{\prime}\right)
-\left(  \frac{d\mu_{A}\left(  t^{\prime}\right)  }{dt^{\prime}}\right)  ^{2}%
}dt^{\prime}\right]  ^{2}}\text{.} \label{ac3}%
\end{equation}
It is important to highlight that $\left[  \mathrm{SNR}\left(  A\right)
\right]  _{\min}$ in Eq. (\ref{ac3}) denotes the instantaneous temporal
profile of the minimum threshold for the signal-to-noise ratio $\mathrm{SNR}%
\left(  A\right)  $. Ideally, to ensure high signal quality, one would prefer
this lower bound to be as large as possible. However, as indicated in Eq.
(\ref{ac3}), higher values of $\langle v_{A}^{2}\rangle$ are associated with
lower values of $\left[  \mathrm{SNR}\left(  A\right)  \right]  _{\min}$. This
observation leads us to understand that elevated values of $\langle v_{A}%
^{2}\rangle$ can negatively impact signal quality by reducing its
instantaneous lower bound. With this final remark, we end this discussion.

\section{Numerical simulations}

In this appendix, we provide additional details regarding the numerical
simulations conducted in our third illustrative example presented in Section
IV, which pertains to a harmonic oscillator situated within a
finite-dimensional Fock space.

For thoroughness, we note that it is not absolutely essential to differentiate
the truncation of the squeezed coherent state within a finite-dimensional
Hilbert space. While the Fock space is theoretically infinite, the expansion
coefficients $\left\{  c_{n}\right\}  $ of the wavefunction $\left\vert
\Psi\right\rangle =\sum_{n}c_{n}\left\vert n\right\rangle $ must adhere to the
conservation of probability, summing to one (i.e., $\sum_{n}\left\vert
c_{n}\right\vert ^{2}=1$).\textbf{\ }Consequently, the coefficients
$\left\vert c_{n}\right\vert ^{2}$ must decrease for sufficiently large $n$ to
ensure that this sum converges to one. In our QuTiP simulation, we select the
mean photon number ($\langle\hat{N}\rangle\overset{\left\vert \alpha
\right\vert ^{2}\ll s}{\approx}\left\vert \alpha\right\vert ^{2}=5$) to be
significantly lower than the dimension of the truncated Fock space ($s=20$,
where we recall that $s+1$ denotes the dimension of the finite-dimensional
Hilbert space generated by the number states\textbf{ }$\left\{  \left\vert
0\right\rangle \text{,..., }\left\vert s\right\rangle \right\}  $). When
$s\gg\left\vert \alpha\right\vert ^{2}$, the truncated coherent state becomes
an excellent approximation of the exact coherent state. Indeed, focusing on
the expectation value of the number operator with respect to the exact and
truncated coherent states, one has $\left\langle \hat{N}\right\rangle
_{\mathrm{exact}}=\left\vert \alpha\right\vert ^{2}$ and,%
\begin{equation}
\left\langle \hat{N}\right\rangle _{\mathrm{truncated}}\overset{\text{def}}%
{=}\frac{%
{\displaystyle\sum\limits_{n=0}^{s}}
n\frac{\left\vert \alpha\right\vert ^{2n}}{n!}}{%
{\displaystyle\sum\limits_{n=0}^{s}}
\frac{\left\vert \alpha\right\vert ^{2n}}{n!}}\text{,} \label{tronco}%
\end{equation}
respectively. We note that $\left\langle \hat{N}\right\rangle
_{\mathrm{truncated}}$ in Eq. (\ref{tronco}) is the weighted average photon
number in the truncated photon number (Poisson) distribution that
characterizes coherent states. A simple calculation shows that if we
pick\textbf{ }$\left\vert \alpha\right\vert ^{2}=5$, then $\left\vert
\left\langle \hat{N}\right\rangle _{\mathrm{exact}}-\left\langle \hat
{N}\right\rangle _{\mathrm{truncated}}\right\vert \simeq10^{-6}$ for a choice
of\textbf{ }$s=20$. In general, to keep the truncation error $\varepsilon$
within a given tolerance threshold, the general guideline is setting
$s\gtrsim\left\vert \alpha\right\vert ^{2}+5\sqrt{\left\vert \alpha\right\vert
^{2}}$ with\textbf{ }$\left\vert \left\langle \hat{N}\right\rangle
_{\mathrm{exact}}-\left\langle \hat{N}\right\rangle _{\mathrm{truncated}%
}\right\vert \leq\varepsilon$. Therefore, experts in quantum optics do not
typically concern themselves with this finite-dimensional Hilbert space
approximation. It is always possible to ensure that the omitted terms
contribute less than a specified tolerance epsilon by including additional
terms in the sum that defines the wavefunction's expansion. With this final
comment, we conclude our discussion regarding the nature of the numerical
simulations utilized in this study.

\bigskip

\bigskip

\end{document}